\documentclass[10pt,pre,aps,twocolumn,superscriptaddress,floatfix,notitlepage]{revtex4-2}
\usepackage[latin1]{inputenc}
\usepackage{hyperref}
\hypersetup{
    colorlinks=true,
    linkcolor=blue,
    filecolor=magenta,      
    urlcolor=cyan,
    pdftitle={Quantum Measurement Induced Radiative Processes in Continuously Monitored Optical Environments},
    pdfpagemode=FullScreen,
    }
\usepackage{mathrsfs,dsfont}
\usepackage{amsmath}
\usepackage{amsfonts,natbib}
\usepackage{amssymb}
\usepackage{bm}
\usepackage[pdftex]{graphicx}

\usepackage{graphicx}
\begin{document}
\title{Quantum measurement induced radiative processes in continuously monitored optical environments}
\author{Eldhose Benny}
\affiliation{Department of Physics, National Institute of Technology Calicut, Kozhikode, Kerala 673601, India}
\author{Sreenath K. Manikandan}
\email{sreenath.k.manikandan@su.se}
\affiliation{Nordita, Stockholm University and KTH Royal Institute of Technology, Hannes Alfv\'{e}ns v\"{a}g 12, SE-106 91 Stockholm, Sweden}
\date{\today}

\begin{abstract}
We investigate the emission characteristics of a measurement-driven quantum emitter in a continuously monitored optical environment. The quantum emitter is stimulated by observing the Pauli spin along its transition dipole that maximally noncommutes with the Hamiltonian of the emitter. It also exchanges energy resonantly with the optical environment, observable as quantum jumps corresponding to the absorption or emission of a photon and the null events where the quantum emitter did not make a jump. 
We characterize the finite-time statistics of quantum jumps and estimate their covariance and precision using the large deviation principle.   While the statistics of absorption and emission events are generically sub-Poissonian with an improved precision by at most a factor of two compared to Poissonian jumps, our analysis also reveals a spin-measurement-induced transition from super-Poissonian to sub-Poissonian in their sum. 
 We conclude by describing generalized quantum measurement strategies using mode-entangled optical beams to access the predicted counting statistics in experiments, with implications extending to optimal quantum clocks.     
\end{abstract}
	\maketitle
 \section{Introduction}
 Quantum emitters which can produce single photons with nonclassical features on demand, as well as photodetectors which can probe and verify their nonclassicality are essential elements of the quantum photonics toolkit for quantum technologies~\cite{Novotny_Hecht_2012,Mandel_Wolf_1995,fortsch_versatile_2013,yang_deterministic_2024,castillo-moreno_dynamical_2024}. Observing sub-Poissonian statistics
 is particularly exciting in these settings, as it cannot be explained within a classical framework~\cite{mandel_sub-poissonian_1979,short_observation_1983,subP}. Null detection events, where the detector does not click within the measurement window, also offer an exciting problem of fundamental interest, as the null detections also carry the valuable information that the quantum emitter is less likely to be in its excited state from which the emission process occurs. Hence, they induce a statistically irreversible evolution towards the quantum ground state of the emitter even when no detections are registered. Null measurements are also of great significance in the context of the quantum measurement problem~\cite{jordan_quantum_2024}, as they belong to the class of quantum measurements that can be reversed or undone in experiments with sequential measurements, with some probability of success~\cite{JordanUndoing,jordan_uncollapsing_2010}.
  
Good emitters are also good detectors, and characterizing their spontaneous emission process, which also lacks classical reasoning, is fundamental to both quantum emitters and detectors alike~\cite{manikandan_detecting_2024}. Its controlled manipulations, including reversibility, are topics of great contemporary interest~\cite{purcell1995spontaneous,agarwal_quantum_1974,lewalle_measuring_2020,jordan_anatomy_2016,minev_catch_2019,PhysRevA.99.022117,kannan_waveguide_2020}. Spontaneous emission is caused by the quantum emitter interacting with vacuum modes of the radiation field, and can be understood as driven by quantum fluctuations of the vacuum. The quantum emitter can also be coherently driven, and the resulting resonant fluorescence emission can be monitored time-continuously via different detection schemes, photon counting, homodyne, or heterodyne detection of the emitted photon~\cite{lewalle_measuring_2020,jordan_anatomy_2016,PhysRevA.62.023805}. Such detection schemes are currently accessible experimentally in artificial atoms~\cite{PhysRevX.6.011002,ficheux_dynamics_2018,PhysRevX.13.021039} where they also find various applications in readout and feedback control~\cite{PhysRevLett.133.153602,PhysRevLett.117.060502,PhysRevX.11.031045}. 

 \begin{figure}[b]
\includegraphics[width=\linewidth]{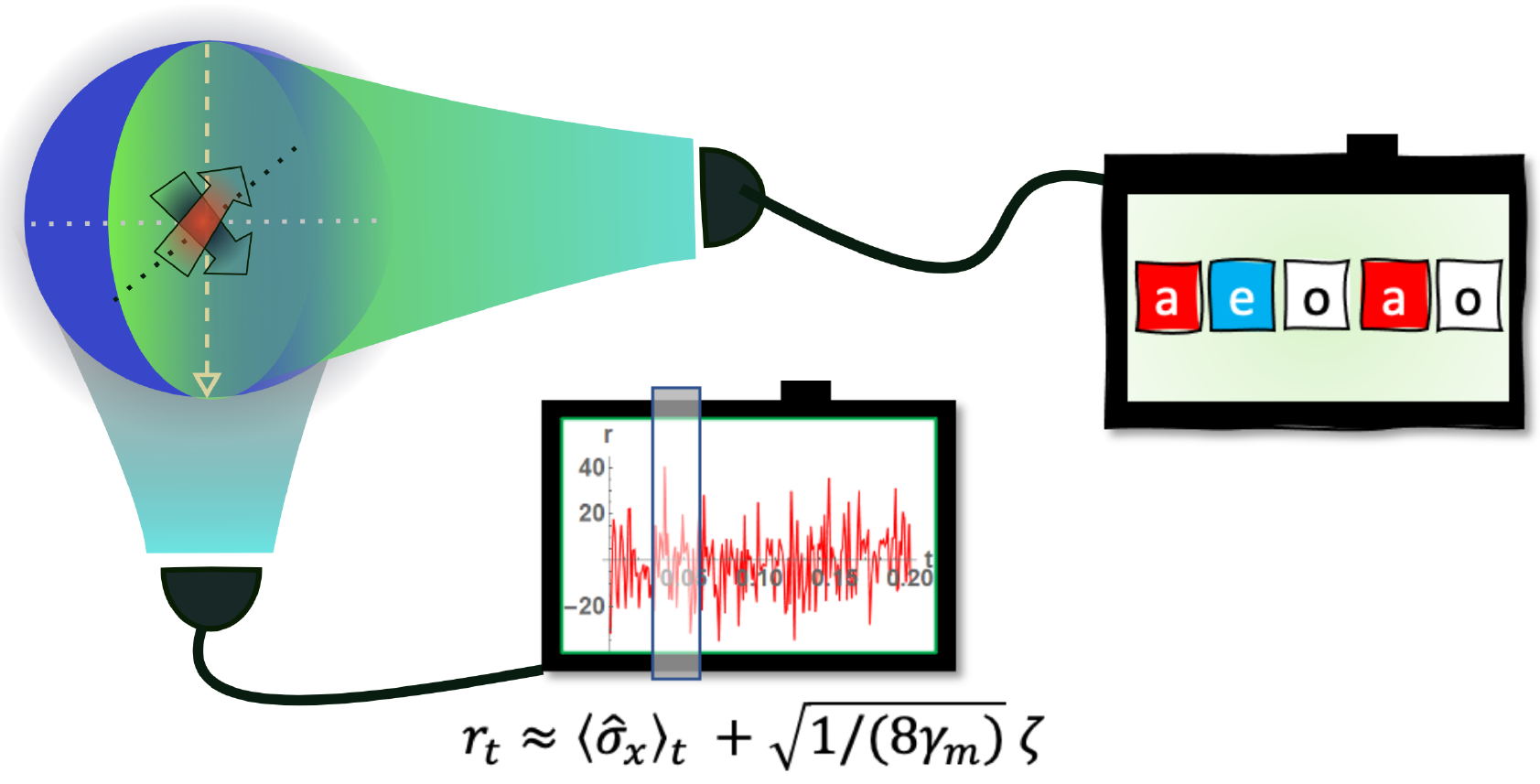}
    \caption{A quantum emitter driven by the quantum measurements of the Pauli spin $\hat{\sigma}_x$ along its transition dipole moment. The induced radiative processes [absorption (a), emission (e), and null events(o)] are recorded by continuously monitoring the optical environment.\label{Figintro}}
\end{figure}

We consider alternate paradigms where quantum fluctuations of different origins drive the radiative processes of a quantum emitter. One source of these fluctuations we consider is the time-continuous quantum measurements of the Pauli spin $\hat{\sigma}_x$ along the quantum emitter's transition dipole, a spin direction that maximally does not commute with the Hamiltonian of the quantum emitter (see Fig.~\ref{Figintro}). This was proposed recently by one of the authors in Ref.~\cite{manikandan_clocks} as a novel strategy to stimulate the quantum emitter incoherently, while still producing coherent features in the subsequent emission.   

The second source of stimulation is the optical environment (the photodetector) itself, which is also a key novelty in our present work~\footnote{The optical environment can also be thought of as resonant photodetector in the single-mode approximation for the optical field, a framework typically used in the description of continuous measurements~\cite{gross_qubit_2018,jordan_quantum_2024}.}. We consider scenarios where the quantum emitter's radiative processes are monitored using optical environments that are initially populated.  This generalizes the standard scenario in optical photodetection, where the environment is reset rapidly to the vacuum state.   
 In the generic scenarios within the rotating wave approximation that we consider in the present article, initially populated photodetectors can either register clicks by absorbing or emitting a photon, or not produce a click (a null event) conditioned on the quantum state of the emitter.  The quantum emitter effectively acts as a transducer in this setting, converting the fluctuations caused by quantum-noise-limited measurements to produce a series of events of varying quantum statistics. 
 
Although the dynamics of the quantum emitter in such generalized optical environments is still straightforward---it either makes a jump by absorbing or emitting a photon, or does not make a jump---identifying appropriate quantum measurement strategies that allow one to track the associated quantum jump statistics in generic optical environments is nontrivial. This is because the above description---jumps or no jump---corresponds to a minimalistic (and optimally informative) unraveling of the quantum emitter's dynamics in the operator-sum representation, which cannot be resolved easily in generic optical environments. However, optical environments rapidly resetting to an excited Fock (number) state make this resolution straightforward, which also demonstrates an improvement in the observable precision of quantum jumps by at most a factor of two when compared to Poissonian jumps.  But the improved precision in quantum jumps when the quantum emitter is probed by an environment in a Fock state is not very surprising, as photons in Fock states are maximally antibunched (Mandel's $Q=-1$), which leads to more regular quantum jumps made by the emitter. This motivates us to further explore if there are other generalized quantum measurement strategies that allow us to access the statistics we propose in experiments. We address this challenge by identifying coarse-grained quantum measurements using two-mode entangled light beams that both generalize our findings as well as broaden the scope of observing the statistical features we predict in experiments.   

Although not in the same spirit as we consider, spontaneous emission in generic environments (such as in squeezed vacuum state~\cite{PhysRevLett.58.2539}), has been investigated in experiments~\cite{PhysRevX.6.031004}.  Such generic environments also offer a much richer scenario without the rotating wave approximation~\cite{PhysRevLett.58.2539,PhysRevX.6.031004}, which we limit ourselves to in the present work for simplicity. Our work is naturally generalizable.

\section{The dynamics}\label{LD_approach}
We consider a continuously monitored quantum emitter interacting with a photodetector that is initially populated. The two-level quantum emitter has energies $0$ and $\Omega$, with the free Hamiltonian of the quantum emitter given by $\hat{H}_0=\Omega |e\rangle\langle e|$. For electronic or generic charge ($q$) transitions between the ground state $|g\rangle\equiv |\psi_g\rangle$ and the excited state $|e\rangle\equiv |\psi_e\rangle$, we can define a transition dipole operator (that couples to external electric fields or coherent drives) as,
\begin{equation}
    \hat{\textbf{d}} = q\langle\psi_e| \hat{\textbf{r}}|\psi_g\rangle |\psi_e\rangle\langle \psi_g|+q\langle\psi_g| \hat{\textbf{r}}|\psi_e\rangle |\psi_g\rangle\langle \psi_e|. 
\end{equation}
For real electronic or charge wavefunctions $\psi_{e(g)}(\textbf{r})$, $\langle\psi_e| \hat{\textbf{r}}|\psi_g\rangle = \langle\psi_g| \hat{\textbf{r}}|\psi_e\rangle$. Hence, we can write,
\begin{equation}
    \hat{\textbf{d}} = d\hat{\sigma}_x, ~~\text{where}~~ d = q\langle\psi_g| \hat{\textbf{r}}|\psi_e\rangle \in R,
\end{equation}
and $\hat{\sigma}_x = |e\rangle\langle g|+|g\rangle\langle e|$ is the Pauli spin observable along the spin-$x$ direction.  The transition dipole moment $d\neq 0$ as long as the dipole transition is not forbidden. 
Instead of applying a coherent drive to stimulate the emitter, we now consider stimulating the quantum emitter's radiative processes by continuously observing its transition dipole operator, which does not commute with the Hamiltonian of the quantum emitter. Since the transition dipole operator is proportional to $\hat{\sigma}_x$, observing the quantum emitter's Pauli spin $\hat{\sigma}_x$ is equivalent to monitoring the transition dipole itself. We can use the following spin measurement operators~\cite{jacobs_straightforward_2006},
\begin{equation}
    \hat{M}(r_t)=(4\epsilon_{m})^{1/4}\exp[-2\epsilon_{m}(r_t\hat{\mathbb{I}}-\hat{\sigma}_{x})^{2}]/\pi^{1/4},
\end{equation}
where $\epsilon_{m}=\gamma_{m}dt$, $\gamma_m$ is the continuous spin measurement rate, and $dt$ is the time-interval between two successive measurements. The time-continuous readout signal $r_t$ at any given time can be represented as $r_t = \langle \sigma_x\rangle_t+\sqrt{1/(8\gamma_m)}\zeta$, where $\zeta$ is a delta-correlated white noise process with variance $1/dt$. The mapping to transition dipole also suggests a natural way to implement the Pauli spin $\hat{\sigma}_x$ measurement by engineering quantum measurement interactions to the transition dipole of the quantum emitter (or to the position $\hat{\textbf{r}}$ of the electron or charged particle in the unharmonic potential), however, note that Pauli spin measurements can be implemented in practice in different ways which are all mathematically equivalent. We also leverage this flexibility, as will be shown briefly, we do not need to keep track of the measurement outcomes in order to stimulate the quantum emitter. It is sufficient to engineer spin dephasing along the transition dipole, corresponding to unconditional spin measurement dynamics of the Pauli spin $\hat{\sigma}_x$.

 We now proceed to the details of observing the radiative processes of the quantum emitter. Starting from the interaction Hamiltonian,
\begin{equation}
    \hat{H}_{int}dt=\sqrt{\gamma_{w} dt}(\hat{\sigma}_{+}\hat{a}+\hat{\sigma}_{-}\hat{a}^{\dag}),
\end{equation}
we obtain the following measurement operators to leading order corresponding to the radiative events (see the Appendixes~\ref{appendix_A} and~\ref{Appendix_B}),
\begin{equation}
    \hat{M}_{w}(e)\approx-i\begin{pmatrix}
                     0  & 0\\
                    \sqrt{(\langle n\rangle+1)\epsilon_w} & 0
                    \end{pmatrix},\label{eqMe} 
\end{equation}    
\begin{equation}
            \hat{M}_{w}(a)\approx-i\begin{pmatrix}
                  0  & \sqrt{\langle n\rangle  \epsilon_w}\\
                  0  & 0
            \end{pmatrix},  
\end{equation}
\begin{equation}
     \hat{M}_{w}(o)\approx\begin{pmatrix}
            \sqrt{1-(\langle n\rangle+1)\epsilon_w} & 0\\
            0           & \sqrt{1-\langle n\rangle\epsilon_w}
            \end{pmatrix}, \label{EqMo}   
\end{equation}
where $\epsilon_w = \gamma_w dt$, $\gamma_w$ is the spontaneous emission rate and $\langle n\rangle$ is the population of the excited photodetector. The labels $a,e, o$ correspond to the absorption, emission, and null events, respectively. 

The conditional time evolution of the quantum emitter for an infinitesimal time interval $dt$ is given by,
\begin{equation}
\rho(t+dt)=\frac{\hat{\mathcal{M}}_{dt}(r_t,k)\rho(t)\hat{\mathcal{M}}^{\dagger}_{dt}(r_t,k)}{\text{tr}\{\hat{\mathcal{M}}_{dt}(r_t,k)\rho(t)\hat{\mathcal{M}}^{\dagger}_{dt}(r_t,k)\}},
\end{equation}
where $\hat{\mathcal{M}}_{dt}(r_t,k) = e^{-i\hat{H}_0 dt}\hat{M}_w(k)\hat{M}(r_t)$, where $-\infty <r_t<\infty$ is the continuous spin readout, and $k\in\{a,e,o\}$ represent the observed radiative process. Note that the quantum state of the emitter evolves even for the null result. Going forward, since we are only interested in the statistics of observable radiative processes, we will limit (including in our simulations) to semi-conditional dynamics that average over different realizations of the spin measurement outcomes $r_t$. This will have an effect similar to adding a dephasing noise, along the $\hat{\sigma}_x$ direction.  

The unconditional dynamics is obtained by averaging over the measurement outcomes of both the spin and radiative processes,
\begin{equation}
\rho(t+dt)=\sum_k\int dr \hat{\mathcal{M}}_{dt}(r_t,k)\rho(t)\hat{\mathcal{M}}^{\dagger}_{dt}(r_t,k).
\end{equation}

To small $dt$,
one obtains the unconditional master equation of the following form (note that $\rho$ now represents the unconditional density matrix, however, we use the same symbol for simplicity),
\begin{eqnarray}
    \frac{d\rho}{dt} &=&-i[\hat{H}_0,\rho]-\gamma_m[\hat{\sigma}_x,[\hat{\sigma}_x,\rho]]\nonumber\\&+&\gamma_w (\langle n\rangle+1)[\hat{\sigma}_{-}\rho\hat{\sigma}_{+}-(\hat{\sigma}_{+}\hat{\sigma}_{-}\rho+\rho\hat{\sigma}_{+}\hat{\sigma}_{-})/2]\nonumber\\
    &+&\gamma_w \langle n\rangle[\hat{\sigma}_{+}\rho\hat{\sigma}_{-}-(\hat{\sigma}_{-}\hat{\sigma}_{+}\rho+\rho\hat{\sigma}_{-}\hat{\sigma}_{+})/2].\label{lindblad}
\end{eqnarray}
The steady state of the unconditional dynamics is given by $\rho_s = \text{diag}\{p,1-p\}$, where
\begin{equation}
    p = \frac{2\gamma_m +\gamma_w\langle n\rangle }{4\gamma_m +\gamma_w(2\langle n\rangle +1)}.
\end{equation}

The only assumption made in arriving at Eq.~\eqref{lindblad} is that the photodetector, or the optical field, is rapidly refreshing to a quantum state of the same average occupation, $\langle n\rangle$. The Gorini-Kossakowski-Sudarshan-Lindblad form of the master equation indeed suggests that the result should hold for a thermal state of the environment with $\langle n\rangle \rightarrow\bar{n}$, the average thermal population of the environment. 

  In many regards, what we arrive at in Eqs.~\eqref{eqMe}---\eqref{EqMo} is a minimalistic unraveling for the thermalization of a quantum emitter in the operator-sum representation that is also maximally informative. `Minimalistic' in that it only uses three measurement operators, and `maximally informative' because it directly corresponds to the jumps or no-jump cases possible. However, experiments that allow one to access this optimal unraveling can be more challenging, as a simple thermal detector will not allow us to resolve the quantum jumps optimally with the few measurement operators that we have. To this end, we describe generalized quantum measurement strategies using mode-entangled light beams in Sec.~\ref{expt} as well as Appendix~\ref{Appendix_B} as one possibility to implement the unraveling we suggest using measurement operators from Eqs.~\eqref{eqMe}---\eqref{EqMo} in experiments.

It is also true that, assuming rapid resetting of the optical environment, the operator-sum representation of the Lindblad dynamics using measurement operators in Eqs.~\eqref{eqMe}---\eqref{EqMo} can be associated to any arbitrary state of the environment with $\langle n\rangle$ being the average population of the environment, provided we make the rotating wave approximation for the interaction between the quantum emitter and its environment (see Appendix.~\ref{appendix_A}). In what follows, we will assume this generality, however, note that they need not all be accessible in experiments in a straightforward manner. Generalizing our considerations beyond the rotating wave approximation is also certainly interesting, as it can probe interesting quantum correlations in the environment~\cite{PhysRevX.6.031004}.  

To compute the observable statistics numerically, we simulate the 
conditional dynamics of the quantum emitter. However, their statistics can be approximated analytically in closed form using the principle of large deviation.  Large deviation theory and comparable approaches have made important contributions to our current understanding of quantum transport in nanoscale quantum devices and solid-state qubits~\cite{PhysRevLett.90.206801,sukhorukov_conditional_2007,PhysRevB.67.085316,PRXQuantum.5.020201}. This technique has been applied to compute the statistics of resonance fluorescence (with the photodetector in the vacuum state) from coherently driven atoms~\cite{garrahan_thermodynamics_2010}, and measurement-driven artificial atoms~\cite{manikandan_clocks}.   In our generalized setting, we obtain the relevant moment-generating function $\theta(s_a,s_e)$ 
as the largest real eigenvalue of the following tilted Lindblad generator,
\begin{equation}
\begin{split}
    W(s_a,s_e)[\rho] =&-i[\hat{H}_0,\rho] - \gamma_{m} [\hat{\sigma}_{x},[\hat{\sigma}_{x},\rho]]\\
    +&\gamma_{w}(\langle n\rangle+1)(e^{-s_{e}}\hat{\sigma}_{-}\rho\hat{\sigma}_{+} 
    -\frac{1}{2}\{\hat{\sigma}_{+}\hat{\sigma}_{-},\rho\})\\
    +&\gamma_{w}\langle n\rangle(e^{-s_{a}}\hat{\sigma}_{+}\rho\hat{\sigma}_{-}-\frac{1}{2}\{\hat{\sigma}_{-}\hat{\sigma}_{+},\rho\}),
\end{split}\label{lb}
\end{equation}
where the $s_a$ and $s_e$ are the tilt parameters. We obtain the following result for the moment generating function,

\begin{eqnarray}
     \theta(s_a,s_e)&=&\frac{1}{2}\big[-(4\gamma_{m}+\gamma_{w}+2\gamma_{w}\langle n\rangle)+
e^{-(s_{a}+s_{e})}\sqrt{g}\big],\nonumber\\\label{mog}
\end{eqnarray}
where,
\begin{eqnarray}
g&=&e^{s_{a}+s_{e}}\big\{4\gamma_{w}(\langle n\rangle+1)(2e^{s_{a}}\gamma_{m}+\gamma_{w}\langle n\rangle) \nonumber\\&+& e^{s_{e}}\left[e^{s_{a}}(16\gamma_{m}^{2}+\gamma_{w}^{2})+8\gamma_{m}\gamma_{w}\langle n\rangle\right]\big\}
\end{eqnarray}

Compared to Ref.~\cite{manikandan_clocks}, which is recovered in the limit $\langle n\rangle\rightarrow 0$, the moment generating function given above is more general, allowing to predict the statistics of the absorption, emission events, and their covariances as shown below. 

\section{Results}
The moments of observed radiative events can be estimated from the large-deviation-approximate form of the moment-generating function given in Eq.~\eqref{mog}. The average and variance of photons absorbed (emitted) within a finite time interval $t$ can be determined as,
\begin{equation}
\begin{split}
   \langle{N_{a(e)}}\rangle=-& \partial_{s_{a(e)}} \theta(s_a,s_e)\big|_{s_a,s_e=0}t \\
   =&\frac{\gamma_w x_{a(e)}(2\gamma_m+\gamma_w x_{e(a)})}{4\gamma_m+2\langle n\rangle\gamma_w+\gamma_w}t,
\end{split}
\end{equation}
and,
\begin{equation}
\begin{split}
    \langle\Delta N_{a(e)}^2\rangle=&\partial_{s_{a(e)}}^{2}\theta(s_a,s_e)\big|_{s_a,s_e=0}t \\
    =&\langle{N_{a(e)}}\rangle -\frac{2\gamma_w^{2}x_{a(e)}^{2}(2\gamma_m+\gamma_w x_{e(a)})^{2}t}{(4\gamma_m+2\langle n\rangle\gamma_w+\gamma_w)^{3}}.\\
\end{split}
\end{equation}
where $x_{a}=\langle n\rangle$ and $x_{e}=\langle n\rangle+1$.
Given the expression of variance, it is evident that it is consistently less than the mean, indicating a sub-Poissonian photon distribution for both the absorption and emission events separately. 
 
Trying to compute the statistics of null events directly can be pathological as it depends on the discretization in time that seems arbitrary (although waiting time-distributions can be computed and are statistically meaningful~\cite{jordan_uncollapsing_2010}). Instead, we make statistically meaningful inferences about the null events from the joint statistics of absorption and emission events, as together they are complete.   
The average number of total jumps in this case is,
\begin{equation}
    \langle \tilde{N}_{o}\rangle = \langle N_{e}\rangle+\langle N_{a}\rangle,
\end{equation}
which can be thought of as the statistical observable for the negation of null events (hence the subscript $``o"$).
The variance of the sum of absorption and emission events can be computed as,
\begin{equation}
    \langle\Delta \tilde{N}_{o}^2\rangle=\langle\Delta N_{e}^2\rangle+\langle\Delta N_{a}^2\rangle+2\text{Cov}(N_e,N_a),\label{Ncov}
\end{equation}
where $\text{Cov}(N_e, N_a)=\partial_{s_as_e}^{2}\theta(s_a,s_e)\big|_{s_a,s_e=0} t$ is the covariance of the observed absorption and emission events, given by,

\begin{eqnarray}
&&\text{Cov}(N_e,N_a)= \nonumber\\&&\frac{\gamma_w^2 x_a x_e \big\{8\gamma_m^2+4\gamma_m (2\gamma_w \langle n\rangle+\gamma_w)+\gamma_w^2 \left[2 x_ax_e+1\right]\big\}}{(4\gamma_m+2\gamma_w \langle n\rangle+\gamma_w)^{3}}t.\nonumber\\
\end{eqnarray}

We use the Mandel's Q parameter, defined as~\cite{mandel_sub-poissonian_1979,short_observation_1983},
 \begin{equation}\label{eqn:Mandel's Q}
     Q=\frac{\langle\Delta N^2\rangle-\langle{N\rangle}}{\langle{N}\rangle},
 \end{equation}
 to quantify the quantumness of observed events. 
 Mandel's Q ranges between $-1<Q<0$ for sub-Poissonian statistics,  $Q=0$ for Poissonian statistics, and $Q>0$ for super-Poissonian statistics. 
For the absorption (or emission) event, we obtain the following Mandel's Q,
\begin{equation}
\begin{split}
Q_{a(e)}&=-\frac{\partial_{s_{a(e)}}^{2}\theta(s_a,s_e)\big|_{s_a,s_e=0}}{\partial_{s_{a(e)}} \theta(s_a,s_e)\big|_{s_a,s_e=0}}-1 
    \\&=-\frac{2 \gamma_w x_{a(e)} \left(2\gamma_m+\gamma_w x_{e(a)}\right)}{(4 \gamma_m+2\langle n\rangle\gamma_w +\gamma_w)^2}, 
\end{split}
\end{equation}
 As discussed, with the rapid refreshing of the detector, the observable statistics are sub-Poissonian for both the absorption and emission events. 
The Mandel's Q parameter $\tilde{Q}_o$ can be calculated similarly for the sum of absorption and emission events (the negation of null events). We obtain,

\begin{eqnarray}
&&\tilde{Q}_o=\nonumber\\&&\frac{-4\gamma_{m}^{2}\gamma_w+ x_ax_e\gamma_{w}^{3}}{\left(\gamma_m(2\langle n\rangle+1)+x_ax_e\gamma_w\right)\left(4\gamma_m+(2\langle n\rangle+1)\gamma_w\right)^2 }.\label{Qnull}\nonumber\\
\end{eqnarray}

Before we move on to comparisons with numerical results, we wish to point out some limitations to our present approach, partially also informed by comparing our numerical simulations to the theoretical predictions. We observe that much smaller step-size $dt$ as well as a large number of realizations are required to verify our predictions, particularly for the negated null statistics at higher occupations of the field, $\langle n\rangle > 1$. The deviations become more significant and pronounced in the negated null statistics at large $\langle n\rangle$, although the agreement for absorption and emission events appears to be much less sensitive to the choice of $\langle n\rangle$ provided $\gamma_w dt \langle n\rangle \ll 1$. 

Since this warrants some explanation, first, of course, we note that the event count rates are approximately doubled for negated null statistics by counting both absorption and emission events, so this introduces the possibility of accumulation of errors that will have a sensitive dependence on the step-size $dt$ as well as the ensemble size. However, since we observe from the simulations that the deviations from theory at large $\langle n\rangle $ are small for the counting statistics of absorption and emission events individually, the more pronounced deviations in the negated null event statistics at higher field occupations $\langle n\rangle$ appear to be introduced through the covariance of absorption and emission events. Although we directly estimate the variance of the sum observable from simulations, it depends on the covariance through Eq.~\eqref{Ncov} which appears to be very sensitive to ensemble characteristics (the step-size $dt$, and number of realizations), making accurate estimates for the negated null statistics from simulations challenging for large $\langle n\rangle$.

These observed challenges and our choice to restrict the analysis to small $\langle n\rangle$ seem rather justified from theoretical, numerical, and pragmatic perspectives to some extent. Firstly, our theoretical predictions only make use of a leading order estimate for the large deviation form of the exact moment-generating function, which can be potentially improved further~\cite{PhysRevLett.122.130605,carollo_large_2021}.  We also limit our considerations to the linear order in perturbation theory that is generically valid for small $\gamma_w\langle n\rangle dt\ll 1$, corresponding to the exchange of zero or one photon with the optical environment. Therefore, it also seems physically meaningful to restrict to small $\langle n\rangle$ where higher-order processes can be faithfully neglected. The second justification is more pragmatic as shown in Sec.~\ref{prec}. We notice that the precision of quantum jumps and Mandel's Q achievable for $\langle n\rangle\rightarrow 1$ are already very close to the corresponding optimal estimates our theory predicts at large $\langle n\rangle$, suggesting that we do not gain much by going to large $\langle n\rangle$. 

Although it is beyond the scope of our present work, we note that improvements in numerical simulation strategies may also help resolve some of the issues from the accumulation of errors from finite step-size, $dt$. An example would be to use the Gillespie algorithm, which generates quantum jump trajectories based on finite-time waiting distributions, as was recently proposed~\cite{Gillespie}. We defer such considerations to future work.  
  
\subsection{The precision of quantum jump events\label{prec}}  
\begin{figure}[t]
\includegraphics[width=\linewidth]{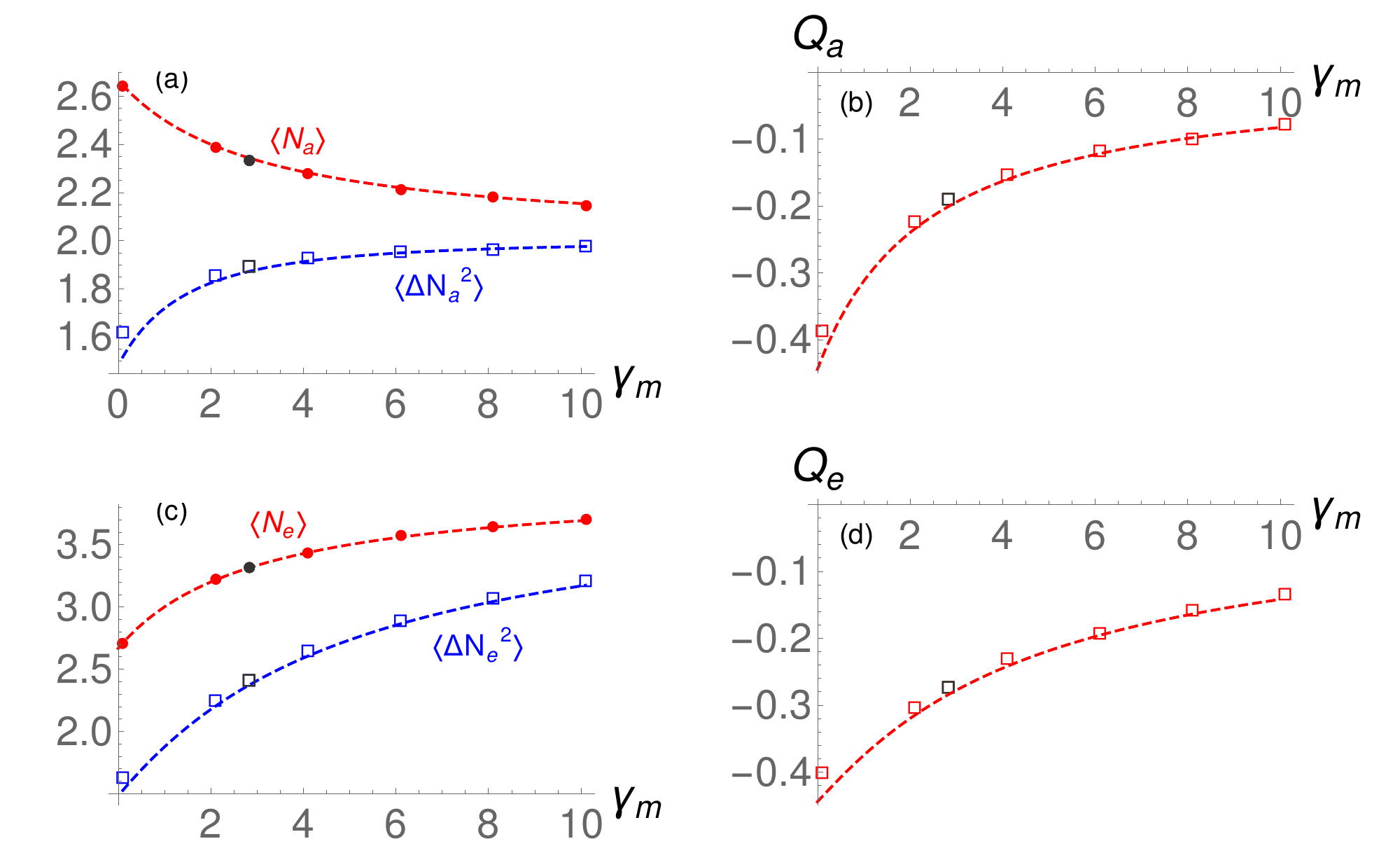}
    \caption{Characterizing the statistics of absorption and emission events: Panels (a) and (c) show the mean and variance of the photons absorbed and emitted, respectively. We use $\Omega=1,\gamma_w=4$ and $\langle n\rangle =1$. In the simulations, the moments are calculated over $8\times10^4$ trajectories, using 2000 time steps with a time increment of $dt=5\times10^{-4}$. Each trajectory is initialized in the steady state $\rho_s$ of the unconditional dynamics. Panels (b) and (d) shows Mandel's Q parameters for absorption and emission events, respectively. The dashed lines are predictions based on the large deviation principle.\label{fig:Absorption & Emission}}
\end{figure}

The statistics of absorption or emission events, when monitored individually, are sub-Poissonian ($-1<Q<0$) as shown in Fig.~\ref{fig:Absorption & Emission}. When $\langle n\rangle\neq 0$, the statistics of absorption and emission events are maximally sub-Poissonian in the absence of continuous quantum measurements of the spin ($\gamma_m = 0$) yielding,
\begin{equation}
    Q_{a(e)}^{\text{min.}}(\langle n\rangle) =-\frac{2\langle n\rangle(\langle n\rangle+1)}{(2\langle n\rangle+1)^2}\rightarrow -4/9~~\text{when $\langle n\rangle\rightarrow 1$.}
\end{equation}
We also note that the asymptotic value of $ Q_{a(e)}^{\text{min.}}(\langle n\rangle)\rightarrow -1/2$ in the limit $\langle n\rangle\gg 1$ is quite close to $Q(1)=-4/9$. This suggests that we do not gain much in terms of Mandel's $Q$ by going to a large $\langle n\rangle$ limit. 

Another good measure of the regularity of the quantum jumps, apart from Mandel's Q parameter, is the relative error, $\Delta N/\langle N\rangle$ (where $\Delta N$ is the standard deviation). Such a measure is applicable, for instance,  if one were to keep track of the number of quantum jumps to measure the time elapsed as in a radio-carbon clock~\cite{milburn_thermodynamics_2020}. If observing either the absorption or the emission events is used as a clock signal, then our findings suggest that such a quantum clock would achieve the following relative error,
\begin{equation}
     \frac{\Delta N_{a(e)}}{\langle N_{a(e)}\rangle }=    \frac{\sqrt{1+Q}}{\sqrt{\langle N_{a(e)}\rangle}}\rightarrow\frac{\sqrt{5}}{3\sqrt{ \langle N_{a(e)}\rangle}}~~\text{when $\langle n\rangle\rightarrow 1$.}
\end{equation}
For large $\langle n\rangle$, $\frac{\Delta N_{a(e)}}{\langle N_{a(e)}\rangle }\rightarrow \frac{1}{\sqrt{ 2\langle N_{a(e)}\rangle}}$ which is again close to the achievable relative error in the limit $\langle n\rangle \rightarrow 1$.  In comparison to a radio-carbon clock which is Poissonian ($Q=0$), using either the absorption or emission in the scenarios we consider as the time-reference can improve the clock's precision [defined as $\langle N_{a(e)}\rangle^2/\Delta N_{a(e)}^2$] by close to a factor of two. Practically again, the limit $\langle n\rangle\rightarrow 1 $ offers an improvement in precision by a factor of $9/5=1.8$, which is rather close to the precision achievable with large $\langle n\rangle$.
  
\subsection{A spin-measurement induced transition in the observable statistics}\label{Results}

\begin{figure}[t]
\includegraphics[width=\linewidth]{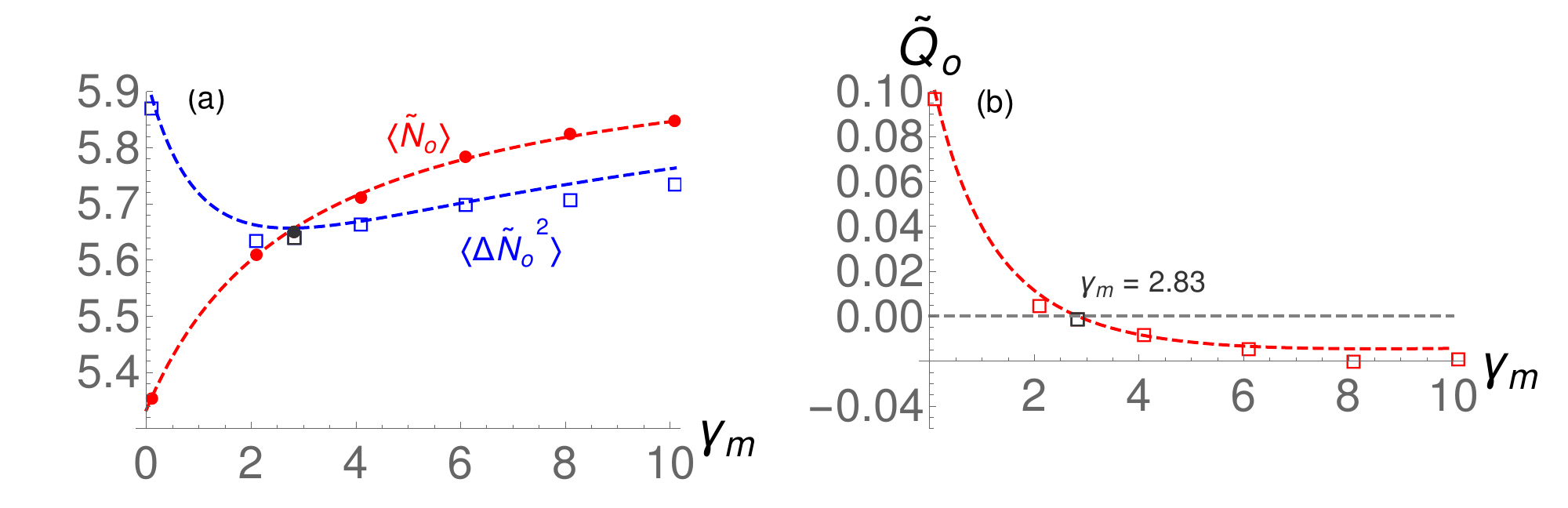}
    \caption{Characterizing the statistics of the sum of absorption and emission events , using the same parameters as in Fig.~\ref{fig:Absorption & Emission}. (a) Mean and variance of the sum of absorption and emission events. (b) The Mandel's Q parameter, with transition point(Q=0) at $\gamma_m=2.83$. The statistics corresponding to the transition point are shown in black. The dashed lines are predictions based on the large deviation principle.\label{fig:null}}
\end{figure}

Keeping track of both the absorption and emission events equally introduces an interesting variability in the observable statistics. From Eq.~\eqref{Qnull}, we note that the corresponding statistics can be sub-Poissonian, Poissonian, or super-Poissonian, depending on the continuous spin measurement rate. In the absence of continuous spin measurements ($\gamma_m\rightarrow 0$),
\begin{equation}
    \tilde{Q}_o \rightarrow (2\langle n\rangle+1)^{-2} > 0,
\end{equation}
suggesting that the observable statistics will be super-Poissonian, approaching Poissonian in the limit of large $\langle n\rangle$. The observable statistics, however, also change as we increase the rate at which the quantum emitter's Pauli spin is continuously monitored along the transition dipole. While we see super-Poissonian statistics ($Q>0$) for low values of $\gamma_m$, the statistics continuously transition into sub-Poissonian statistics by increasing $\gamma_m$ (see Fig.~\ref{fig:null} and Fig.~\ref{fig:null05}.).
The value of $\gamma_m$ at which the transition happens is
\begin{equation}
2\gamma_m=\gamma_w\sqrt{\langle n\rangle^2+\langle n\rangle},
\end{equation}
where the statistics are Poissonian. Rewriting Eq.~\eqref{lb} in the traditional Lindblad form, we can think of $2\gamma_m$ as the rate corresponding to the Lindblad dissipator $ L = \sigma_x$.   Interestingly, the transition point we obtain 
  turns out to be
 the geometric mean of the other two transition rates, $\gamma_w \langle n\rangle$ (corresponding to $L=\hat{\sigma}_+$) and $\gamma_w (\langle n\rangle+1)$ (corresponding to $L=\hat{\sigma}_-$), alluding to the competing nature of various underlying processes that prefer different statistics (thermal driven: super-Poissoian, measurement-driven: sub-Poissonian) involved.  

Our findings here also demonstrate that the covariance of observing the quantum jumps also contains additional extractable information about the quantum nature of the underlying dynamics, not already evident from the statistics of absorption or emission events monitored individually. Physically they can be associated with the distribution of null detection events within a finite-time window of monitoring the quantum jumps.  While it is only a specific quantum emission scenario we consider, it reinforces our understanding that the null detection events in quantum mechanics are not noninformative. The negation of their finite-time statistics we estimate can also be relevant in other comparable scenarios, for example, in the avenue of quantum sensing of weak forces where the null events are not rare~\cite{quantumSensing}.
\subsection{Experimental scenarios involving entangled optical beams\label{expt}}
Here we discuss a possible experimental scenario that could reveal the statistics we predict in experiments, as the generalized measurements we describe deviate from conventional photodetection scenarios where the optical environment (or the photo-detector) is initialized in the vacuum state. As already noted, our results are applicable when the optical environment is rapidly reset to a Fock state (see Appendix~\ref{appendix_A}). However, a limitation of this scenario is that it only applies to the case when $\langle n\rangle$ in Eqns.~\eqref{eqMe}--\eqref{EqMo} is a whole number. Experimental scenarios resolving the quantum jumps as optimally as we propose become more challenging when $\langle n\rangle$ is not a whole number. However, we describe one such scheme below (detailed in Appendix~\ref{Appendix_B}) that leverages the fact that a notion of memory of the prior quantum state of the detector is helpful to achieve the minimalistic unraveling of the Lindblad dynamics we wish to achieve. 

To generalize to arbitrary positive $\langle n\rangle$ in practice, we consider the following optical detection scenario involving two mode-entangled light beams in the following quantum state,
\begin{eqnarray}
    |\psi_{FF'}\rangle&=&\frac{1}{\cosh(r)}\sum_n\tanh^n(r)|n\rangle_F|n\rangle_{F'}\nonumber\\&=& \sum_n \sqrt{P_n} |n\rangle_{F} |n\rangle_{F'}.
\end{eqnarray}
Such two-mode squeezed states can be generated, for example, by using a beam-splitter with two single-mode squeezed vacuum states as inputs~\cite{PLK,eberle_stable_2013,hage_demonstrating_2011,lvovsky_squeezed_2015}. In the optical detection scenario we are interested in, one of the two modes ($F$) can be thought of as the probe mode, which serves as the optical environment to the quantum emitter, and the other one serves as a reference. Following the local interaction between the probe field $F$ and the quantum emitter, we envision performing the following global quantum measurement (using subspace projection operators) that compares the occupation of the probe beam relative to the reference,
\begin{eqnarray}
    \Pi_{a}&=&\sum_n |n-1\rangle |n\rangle \langle n-1| \langle n|,\nonumber\\
     \Pi_{e}&=&\sum_n |n+1\rangle |n\rangle \langle n+1| \langle n|,\nonumber\\
      \Pi_{o}&=&\sum_n |n\rangle |n\rangle \langle n| \langle n|.
\end{eqnarray}
We show in Appendix~\ref{Appendix_B} that on the projection and tracing out the field degrees of freedom, the effective measurement operators for the quantum emitter reduce to Eqns.~\eqref{eqMe}--\eqref{EqMo}, with $\langle n\rangle =\sum_n n P_n$ generalized to any positive number.

This alternative measurement strategy can also be relevant for the extraction of optimal quantum clock signals from quantum emitters that are otherwise autonomous. As we already noted, the observable counting statistics achieve improved precision for nonzero $\langle n\rangle$, however, the true identification of the resource behind this quantum enhancement in precision is dependent on the quantum measurement strategy used. When the optical environment is initialized in a Fock state, the predicted improvement can be associated with the perfect antibunching of photons in the Fock state. The alternate scheme described above trades this resourceful nature of Fock states to the quantum entanglement between the probing field and the reference beam, and an optimal quantum measurement strategy that compares the two beams.    
  
\section{Conclusions}
  We characterized the statistics of the radiative processes of a measurement-driven quantum emitter in initially populated optical environments through an optimal unraveling of quantum jumps and no jumps, using the large deviation principle.   The quantum emitter was driven by continuously monitoring the Pauli spin $\hat{\sigma}_x$ along its transition dipole, a direction of spin that maximally noncommutes with the Hamiltonian of the quantum emitter. We find that the statistics of observed quantum jumps (absorption or emission) are generically sub-Poissonian, and the total number of absorption and emission events undergoes a quantum measurement-driven transition from super-Poissonian to sub-Poissonian statistics as a function of the Pauli spin measurement rate. 

  For the optimal unravelings we have considered,  we observe that the precision of quantum jumps is improved by at most a factor of two when compared to Poissonian jumps. Our results here are applicable to quantum clocks, especially
on the readout/tick register part, where we have demonstrated that optimizing the regularity of the clock
ticks beyond conventional photodetectors is possible by using quantum mechanically resourced readout
strategies, for example, using an initially excited tick register, or correlated photons in place of a conventional tick register.

Our quantum measurement-induced stimulation strategies could also be relevant to the paradigm of stimulating atomic and nuclear transitions for optimal quantum clocks~\cite{RevModPhys.87.637}. There have been important recent developments in clever ways to address nuclear transitions, which also shed light on the difficulty in exciting and observing desired atomic or nuclear transitions in general~\cite{zhang_frequency_2024,PhysRevLett.132.182501}. Inducing optical transitions using the generalized quantum measurement strategies we described can be significant in these contexts, where it may have the potential to stimulate transitions that are difficult 
 to achieve in practice by using traditional alternatives such as applying coherent drives.  There have been some early suggestions on the role of the quantum Zeno effect in nuclear transitions~\cite{panov_quantum_1996} to which the generalized emission scenarios we consider are broadly relatable.   
 
 Our proposals are certainly accessible in artificial atom-clock scenarios~\cite{manikandan_clocks,prasannaClock,milburn_thermodynamics_2020,ErkerClocks}, where they can be used as analog systems to further explore the impact of quantum measurement engineering in real atomic and nuclear transitions.
 It also applies to the paradigm of flying qubits and free-space quantum optics for optimal atom-photon interactions~\cite{PhysRevA.94.033832,stobinska_perfect_2009}.  Our findings can be quite relevant for quantum thermodynamics too, as we propose an effective strategy to mimic the Lindblad dynamics using mode-entangled photon pairs, while also providing a minimalistic resolution to the quantized energy exchanges involved that are physically meaningful. 
 Our results are generalizable to the case of driven atoms and to driven atoms that are continuously monitored. They encompass various avenues of contemporary interest in quantum thermodynamics~\cite{cangemi_quantum_2023,manikandan_clocks,ErkerClocks,milburn_thermodynamics_2020,prasannaClock,elouard_role_2017,PRXQuantum.5.020201,ferri-cortes_entropy_2024,PhysRevX.8.031037,sundelin_quantum_2024,guzman_useful_2024}, further expanding the scope of the methodologies and ideas presented in this article.

\noindent\textit{Data availability statement---}The numerical simulations (using Mathematica~\cite{Mathematica}) associated with the article can be accessed in the following \href{https://github.com/sreenathkm92/QuantumEmitter}{ Github repository link}. 

\section{Acknowledgments} 

SKM thanks Andrew Jordan, Supriya Krishnamurthy, Pranay Nayak, and Sreeram PG for discussions on related topics and Andrea Maiani and Igor Pikovski for helpful suggestions. The work of SKM is supported in part by the Swedish Research Council under Contract No. 335-2014-7424 and in part by the Wallenberg Initiative on Networks and Quantum Information (WINQ). This collaboration was made possible through the Nordita Summer Internship Program in Theoretical Physics, 2024. We extend our sincere thanks to Ivan Khaymovich for his valuable support.

\noindent\textit{Author contributions---} SKM conceptualized the work. Both SKM and EB contributed to deriving the large deviation results, analyzing the simulations, and writing the manuscript. 
\begin{widetext}
\appendix
\section{Measurement Operators for Observable Radiative Processes}\label{appendix_A}

We consider the quantum emitter initialized in an arbitrary, pure initial state, $|\Psi_s\rangle=\alpha|e\rangle+\beta|g\rangle$. To derive the measurement operators corresponding to the absorption, emission, and the null events, we first consider a simple scenario where the quantum emitter is coupled to an excited state of the optical field, $|n\rangle$. The state of the combined system in time $dt$ is,
\begin{equation}
 |\Psi_{dt}\rangle=e^{-i\hat{H}_{int}dt}|\Psi_s\rangle|n\rangle,
\end{equation}
where the interaction Hamiltonian we consider satisfies $\hat{H}_{int}dt=\sqrt{\gamma_{w} dt}(\hat{\sigma}_{+}\hat{a}+\hat{\sigma}_{-}\hat{a}^{\dag})$~\cite{gross_qubit_2018}. Here $\gamma_w$ is the spontaneous emission rate of the quantum emitter, $\hat{\sigma}_+$ and $\hat{\sigma}_-$ are the raising and lowering operators of the emitter, and $\hat{a}^\dag$ and $\hat{a}$ are the creation and annihilation operators for the optical field.

\bigskip
Taylor expanding the time-evolution operator $e^{-i\sqrt{\gamma_{w} dt}(\hat{\sigma}_{+}\hat{a}+\hat{\sigma}_{-}\hat{a}^{\dag})}$ till linear order in $dt$, we obtain,
\begin{equation}
    |\Psi_{dt}\rangle\approx\left [\hat{1}-i\sqrt{\gamma_w dt}(\hat{\sigma}_{+}\hat{a}+\hat{\sigma}_{-}\hat{a}^{\dag})-\frac{\gamma_w dt}{2}\left(|e\rangle\langle e| (\hat{a}^{\dag}\hat{a}+1)+|g\rangle\langle g| \hat{a}^{\dag}\hat{a}\right)\right ]|\Psi_s\rangle|n\rangle.
\end{equation}
The measurement operator corresponding to no photon detection (null event) is obtained by projecting to $|n\rangle$,
\begin{equation}
\begin{split}
    \hat{M}_{w}(o)&=\langle n|\Psi_{dt}\rangle\\
    &=\begin{pmatrix}
            1-\frac{\gamma_w dt (n+1)}{2} & 0\\
            0           & 1-\frac{\gamma_w dt n}{2}
            \end{pmatrix}\\
        &\approx\begin{pmatrix}
            \sqrt{1-(n+1)\gamma_w dt} & 0\\
            0           & \sqrt{1-n\gamma_w dt}
            \end{pmatrix}.
\end{split}
\end{equation}
Similarly, the measurement operators corresponding to emission and absorption are,
\begin{equation}
    \hat{M}_w(e)=\langle n+1|\Psi_{dt}\rangle=-i\begin{pmatrix}
            0  & 0\\
            \sqrt{(n+1)\gamma_w dt}&0
            \end{pmatrix},
\end{equation}
\begin{equation}
    \hat{M}_w(a)=\langle n-1|\Psi_{dt}\rangle=-i\begin{pmatrix}
            0  & \sqrt{n \gamma_w dt}\\
            0  &0
            \end{pmatrix}. 
\end{equation}
The above results assume that the optical field, or the photodetector, is always reset to a Fock state. We will now consider potential generalizations to the above measurement operators. For an arbitrary initial state of the field, $|\psi_F\rangle,$ we can write $|\Psi_{dt}\rangle$ in matrix form as,
\begin{eqnarray}
|\Psi_{dt}\rangle\approx\begin{pmatrix}
         1-\gamma_w dt(\hat{a}^{\dagger} \hat{a}+1)/2 &&-i\hat{a}\sqrt{\gamma_w dt}\\-i\hat{a}^\dagger\sqrt{\gamma_w dt} &&1-\gamma_w dt(\hat{a}^{\dagger} \hat{a})/2 
     \end{pmatrix}|\Psi_s\rangle|\psi_F\rangle.
\end{eqnarray}

Similarly, for a generic density matrix $\rho_s\otimes \rho_F$, we may write,

\begin{eqnarray}
\rho_{dt}&\approx&\begin{pmatrix}
         1-\gamma_w dt(\hat{a}^\dagger \hat{a}+1)/2 &&-i\hat{a}\sqrt{\gamma_w dt}\\-i\hat{a}^\dagger\sqrt{\gamma_w dt} &&1-\gamma_w dt(\hat{a}^\dagger \hat{a})/2 
     \end{pmatrix}\rho_s\otimes \rho_F \begin{pmatrix}
         1-\gamma_w dt(\hat{a}^\dagger \hat{a}+1)/2 &&i\hat{a}\sqrt{\gamma_w dt}\\i\hat{a}^\dagger\sqrt{\gamma_w dt} &&1-\gamma_w dt(\hat{a}^\dagger \hat{a})/2 
     \end{pmatrix}\nonumber\\
     &=&\hat{\mathcal{M}}_{a}\rho_s\otimes \rho_F\hat{\mathcal{M}}_{a}^{\dagger}+\hat{\mathcal{M}}_{e}\rho_s\otimes \rho_F\hat{\mathcal{M}}_{e}^{\dagger}+\hat{\mathcal{M}}_{o}\rho_s\otimes \rho_F\hat{\mathcal{M}}_{o}^{\dagger},
\end{eqnarray}

where we have identified,

\begin{eqnarray}
    \hat{\mathcal{M}}_e = \begin{pmatrix}
         0&&0\\-i\hat{a}^\dagger\sqrt{\gamma_w dt} &&0    \end{pmatrix},~~ \hat{\mathcal{M}}_a = \begin{pmatrix}
         0 &&-i\hat{a}\sqrt{\gamma_w dt}\\0&&0
     \end{pmatrix},
\end{eqnarray}

and,

\begin{equation}
 \hat{\mathcal{M}}_o = \begin{pmatrix}
         1-\gamma_w dt(\hat{a}^\dagger \hat{a}+1)/2 &&0\\0 &&1-\gamma_w dt(\hat{a}^\dagger \hat{a})/2 
     \end{pmatrix}.
\end{equation}

These operators satisfy the completeness relation, 

\begin{equation}
    \sum_{i\in\{a,e,o\}}\text{tr}_{F}[\rho_F\hat{\mathcal{E}}_i]=\sum_{i\in\{a,e,o\}}\hat{E}_i\approx\hat{1}_{2\times 2},
\end{equation}
where $\hat{\mathcal{E}}_i = \hat{\mathcal{M}}_i^\dagger \hat{\mathcal{M}}_i,$ and the effect operators (or POVM elements) defined as $\hat{E}_i = \text{tr}_{F}[\rho_F\hat{\mathcal{E}}_i].$ We find that,
\begin{eqnarray}
    \hat{E}_e = \begin{pmatrix}
         \text{tr}\{\rho_F (\hat{a}\hat{a}^\dagger)\} \gamma_w dt&&0\\0&&0
    \end{pmatrix} =\begin{pmatrix}
         (\langle n\rangle +1) \gamma_w dt&&0\\0&&0
    \end{pmatrix}, \end{eqnarray}
    
    \begin{eqnarray}
        \hat{E}_a = \begin{pmatrix}
          0&&0\\0&&\text{tr}\{\rho_F (\hat{a}^\dagger \hat{a})\}\gamma_w dt
    \end{pmatrix}=\begin{pmatrix}
          0&&0\\0&&\langle n\rangle\gamma_w dt
    \end{pmatrix},
    \end{eqnarray}

and

\begin{eqnarray}
     \hat{E}_o = \begin{pmatrix}
         1-\text{tr}\{\rho_F (\hat{a}\hat{a}^\dagger)\} \gamma_w dt&&0\\0&&1-\text{tr}\{\rho_F (\hat{a}^\dagger \hat{a})\}\gamma_w dt  
    \end{pmatrix}=\begin{pmatrix}
         1-(\langle n\rangle +1) \gamma_w dt&&0\\0&&1-\langle n\rangle \gamma_w dt 
    \end{pmatrix}.\nonumber\\
\end{eqnarray}
\begin{figure}[t]
\includegraphics[width=\linewidth]{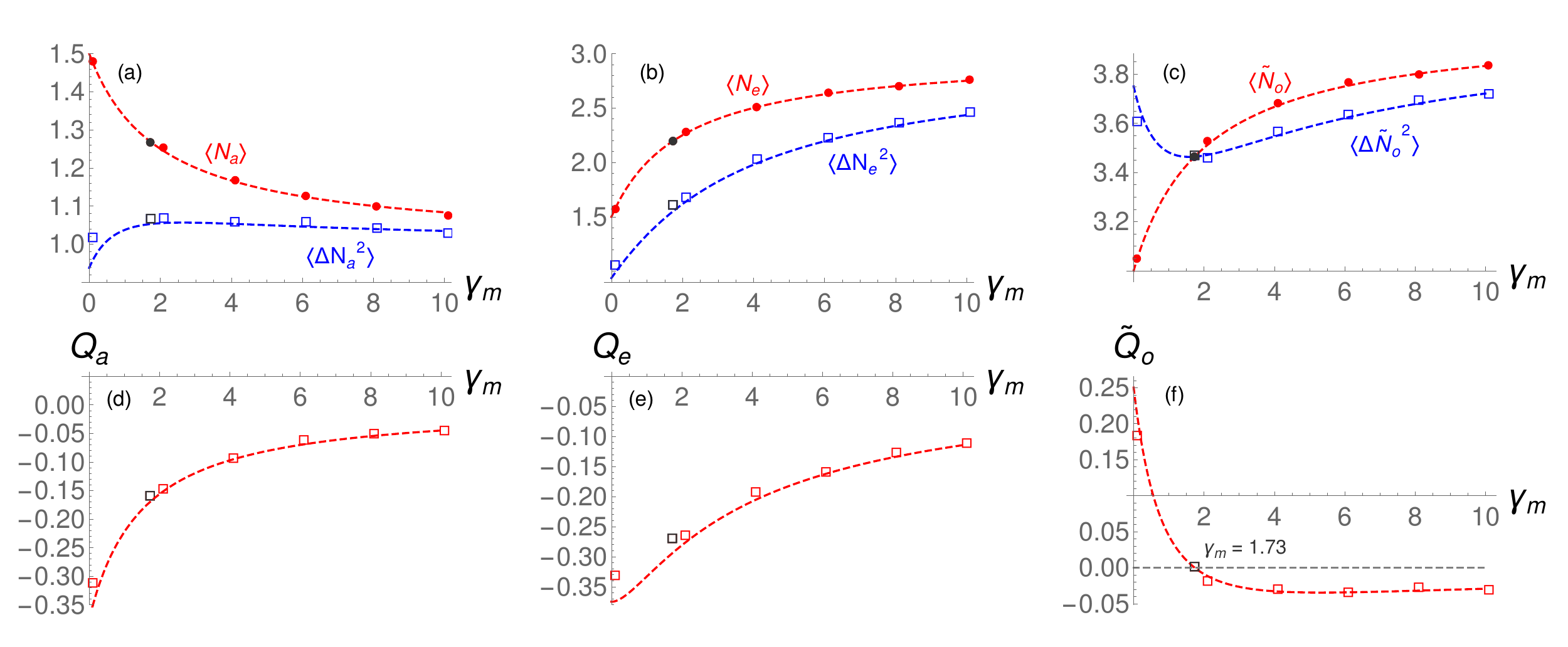} \caption{Statistics of the radiative processes in an optical environment with average occupation number $\langle n \rangle=0.5$: Panels (a) and (b) shows the mean and variance of photons absorbed and emitted respectively and the panels (d) and (e) are the corresponding Mandel's Q parameters. The dashed lines are predictions of the large deviation principle. We assume parameters otherwise identical to Figs.\ref{fig:Absorption & Emission} and \ref{fig:null}. Panels (c) and (f) show the statistics of the negation of null events (sum of absorption and emission events), with the transition from super-Poissonian to sub-Poissonian statistics occurring at $\gamma_m=1.73$. The transition point from simulations is also shown in black.\label{fig:null05}}
\end{figure} 
We have denoted $\text{tr}\{\rho_F \hat{a}^{\dagger}\hat{a}\}=\langle n\rangle,$ the average occupation of the optical environment.  We now resolve the effect operators (or POVM elements) to the following measurement operators, which are similar in form to the ones we found before. However, now they are better understood as the coarse-grained measurement operators that correspond to absorption, emission, or null events in the quantum emitter in arbitrary optical environments,
\begin{equation}
    \hat{M}_e = -i\begin{pmatrix}
            0  & 0\\
            \sqrt{(\langle n\rangle+1)\gamma_w dt}&0
            \end{pmatrix},~~\hat{M}_a = -i\begin{pmatrix}
            0  & \sqrt{\langle n\rangle \gamma_w dt}\\
            0  &0
            \end{pmatrix},\label{mae}
\end{equation}
and,
\begin{eqnarray}
    \hat{M}_o = \begin{pmatrix}
         \sqrt{1-(\langle n\rangle +1) \gamma_w dt}&&0\\0&&\sqrt{1-\langle n\rangle \gamma_w dt}\label{mo}
    \end{pmatrix}.
\end{eqnarray}
Here we have only imposed $\hat{M}_i^\dagger \hat{M}_i = \hat{E}_i$ for $i\in\{a,e,o\}$. Generically, the average occupation $\langle n\rangle$ can be a continuous variable, as opposed to the case when the detector is in a number state. It can be verified that the unconditional evolution generated by these measurement operators to leading order reproduces the Gorini-Kossakowski-Sudarshan-Lindblad form of the master equation given in the main text.   The precise quantum measurement strategy which can lead to the three effective quantum measurement operators described above requires special attention, and we address this in Appendix.~\ref{Appendix_B}.
   Note that we have also limited ourselves to the rotating wave approximated interaction Hamiltonian from the beginning, and therefore interesting quantum effects, such as the effect of a squeezed reservoir if present~\cite{PhysRevLett.58.2539,PhysRevX.6.031004}, do not appear in the dynamics to linear order. However our present approach to determine the statistics can be potentially generalized beyond the rotating wave approximation as well. We defer these considerations to future work.
 
\section{A Quantum Measurement Model for the Emergence of Effective Measurement Operators Using Two Entangled Laser Beams\label{Appendix_B}}
It was shown in Appendix~\ref{appendix_A} that the measurement operators corresponding to absorption, emission, and null events can be derived within the standard quantum measurement model when $n$ is a whole number, assuming that the rapidly refreshing optical environment (which serves as the measurement apparatus in the quantum measurement model) is initialized in a number state $|n\rangle$. The resultant unconditional dynamics is of the Lindblad form, and hence we made the continuation from whole numbers $n\rightarrow \langle n\rangle$  in Eqs.~\eqref{mae}--\eqref{mo}, where $\langle n\rangle $ arbitrary positive real number representing the average occupation of the optical environment. While it was explained in Appendix~\ref{appendix_A} as one possible set of POVM elements that can be derived quite generically, here we present an alternate quantum measurement strategy that derives these measurement operators as resulting from a ``three-outcome'' measurement. In an experiment, this requires that we use two entangled laser beams, one of which interacts with the quantum emitter and the other serves as a reference beam. Creation and applications of two-mode entangled states for various quantum information processing applications using continuous variable states have already been discussed elsewhere~\cite{eberle_stable_2013,hage_demonstrating_2011,lvovsky_squeezed_2015}, one of the straightforward approaches use a beam-splitter with squeezed single mode vacuum states at the inputs to generate output modes that are entangled~\cite{eberle_stable_2013,PLK}. This can in principle, generate entangled laser beams in a two-mode squeezed vacuum state~\cite{lvovsky_squeezed_2015,PLK},
\begin{equation}
    |\psi_{FF'}\rangle=\frac{1}{\cosh(r)}\sum_n\tanh^n(r)|n\rangle_F|n\rangle_{F'}= \sum_n \sqrt{P_n} |n\rangle_{F} |n\rangle_{F'}.
\end{equation}
The state is also known as thermofield double with the mapping $P_n= e^{-\hbar\omega n/(k_{B}T)}/\sum_n  e^{-\hbar\omega n/(k_{B}T)}$ that yields $\langle n\rangle =\sum_n nP_n =\sinh^2(r)$, where r is the squeezing parameter. Here $F$ and $F'$ label the two optical (laser) fields entangled with each other in an experiment, and tracing out one of the field modes reduces the other mode to a thermal state, making the correspondence to an effective Lindblad dynamics locally meaningful.  
We now consider our quantum emitter interacting with one of the two modes, say $F$. The measurements we will perform eventually will also make use of the reference beam $F'$. In what follows, we will drop the $F,F'$ subscripts for simplicity. Considering the same interaction as in Appendix \ref{appendix_A}, the state of the system in time $dt$ is, 
\begin{equation}
    |\Psi_{dt}\rangle\approx\left [\hat{1}-i\sqrt{\gamma_w dt}(\hat{\sigma}_{+}\hat{a}+\hat{\sigma}_{-}\hat{a}^{\dag})-\frac{\gamma_w dt}{2}\left(|e\rangle\langle e| (\hat{a}^{\dag}\hat{a}+1)+|g\rangle\langle g| \hat{a}^{\dag}\hat{a}\right)\right ]|\Psi_s\rangle \otimes \sum_n \sqrt{P_n} |n\rangle |n\rangle,\label{eqnn}
\end{equation}
The interaction makes the quantum emitter absorb or emit a photon, or decay nonradiatively (the null event). Consequently, the occupation of the interacting field can decrease, increase, or remain the same in comparison to the reference beam. Now joint measurements on the interacting mode and the reference beam can be performed to infer the radiative process at every time step. The projection operators corresponding to each of the events are,
\begin{equation}
    \Pi_{a}=\sum_n |n-1\rangle |n\rangle \langle n-1| \langle n|,
\end{equation}
\begin{equation}
    \Pi_{e}=\sum_n |n+1\rangle |n\rangle \langle n+1| \langle n|,
\end{equation}
\begin{equation}
     \Pi_{o}=\sum_n |n\rangle |n\rangle \langle n| \langle n|.
\end{equation}
Note that the completeness of the optical modes can be written as,
\begin{equation}
    \hat{1} = \sum_{nm}|nm\rangle\langle n m| =\Pi_a+\Pi_e+\Pi_0 + (\hat{1}-\Pi_a-\Pi_e-\Pi_0),
\end{equation}
where $\hat{1}-\Pi_a-\Pi_e-\Pi_0$ corresponds to the subspace that is orthogonal to the states accessible from the short-time dynamics described by Eq.~\eqref{eqnn}. We also have, $\langle\Psi_{dt}|(\Pi_a+\Pi_e+\Pi_0)|\Psi_{dt}\rangle = 1$. In that sense, the outcomes $a,e,o$ form a complete set. We can now obtain the density matrix of the system corresponding to each event by projecting onto the respective subspace and tracing over the beams,
\begin{equation}  \rho_{dt}^{k}\propto\text{Tr}_{FF'}\{\Pi_k|\Psi_{dt}\rangle \langle\Psi_{dt}|\Pi_k\} = \hat{M}_k \rho_s \hat{M}^{\dagger}_k,
\end{equation}
where $k\in\{a,e,o\}$ and $\rho_s = |\Psi_s\rangle\langle\Psi_s|$. Now we can identify the measurement operators for each event. For the absorption event, we obtain,
\begin{equation}
\begin{split}
    \hat{M}_a \rho_s \hat{M}^{\dagger}_a &=\text{tr}_{FF'}[\Pi_a|\Psi_{dt}\rangle \langle\Psi_{dt}|\Pi_a]\\
    &=\begin{pmatrix}
         0 &&-i\sqrt{\gamma_w dt}\\ 0 &&0 
     \end{pmatrix}\rho_s\begin{pmatrix}
         0 && 0 \\i\sqrt{\gamma_w dt}&&0 
     \end{pmatrix}\text{tr}_{FF'}\big[\sum_n \sqrt{P_n} \hat{a}|n\rangle|n\rangle \sum_m \sqrt{P_m}\langle m|\langle m|\hat{a}^\dagger\big]\\
     &=\sum_n n P_n \begin{pmatrix}
         0 &&-i\sqrt{\gamma_w dt}\\ 0 &&0 
     \end{pmatrix}\rho_s\begin{pmatrix}
         0 && 0 \\i\sqrt{\gamma_w dt}&&0 
     \end{pmatrix}\\
     &=\begin{pmatrix}
         0 &&-i\sqrt{\langle n\rangle \gamma_w dt}\\ 0 &&0 
     \end{pmatrix}\rho_s\begin{pmatrix}
         0 && 0 \\i\sqrt{\langle n\rangle  \gamma_w dt}&&0 
     \end{pmatrix},
\end{split} 
\end{equation}
where we have defined $\langle n\rangle =\sum_n nP_n$. Similarly, for the emission event, we obtain,
\begin{equation}
\begin{split}
    \hat{M}_e \rho_s \hat{M}^{\dagger}_e &=\text{tr}_{FF'}[\Pi_e|\Psi_{dt}\rangle \langle\Psi_{dt}|\Pi_e]\\
    &=\begin{pmatrix}
         0 &&0\\ -i\sqrt{\gamma_w dt} &&0 
     \end{pmatrix}\rho_s\begin{pmatrix}
         0 && i\sqrt{\gamma_w dt} \\0&&0 
     \end{pmatrix}\text{tr}_{FF'}\big[\sum_n \sqrt{P_n} \hat{a}^\dagger|n\rangle|n\rangle \sum_m \sqrt{P_m}\langle m|\langle m|\hat{a}\big]\\
     &=\sum_n (1+n) P_n \begin{pmatrix}
         0 &&0\\ -i\sqrt{\gamma_w dt} &&0 
     \end{pmatrix}\rho_s\begin{pmatrix}
         0 && i\sqrt{\gamma_w dt} \\0&&0 
     \end{pmatrix}\\
     &=\begin{pmatrix}
         0 &&0\\ -i\sqrt{(1+\langle n\rangle)\gamma_w dt} &&0 
     \end{pmatrix}\rho_s\begin{pmatrix}
         0 && i\sqrt{(1+\langle n\rangle)\gamma_w dt} \\0&&0 
     \end{pmatrix}.
\end{split} 
\end{equation}
The measurement operator for the null event can now be identified as,
\begin{equation}
    \hat{M}_0 = \sqrt{\hat{1}_{2\times 2}-\hat{M}_{a}^\dagger \hat{M}_a-\hat{M}_{e}^\dagger \hat{M}_e}.
\end{equation}
However, we also make a similar calculation as above for null events. We first note that,
\begin{eqnarray}
    \Pi_o |\Psi_{dt}\rangle  &=& \sum_n \bigg[\hat{1}-\frac{\gamma_w dt}{2}\bigg(|e\rangle\langle e|(n+1)+|g\rangle \langle g|n\bigg)\bigg]|\Psi_s\rangle \sqrt{P_n}|n\rangle |n\rangle\nonumber\\
    &\approx&\sum_n \bigg[\sqrt{P_n [1-(n+1)\gamma_w dt]}|e\rangle\langle e|+\sqrt{P_n [1-n\gamma_w dt]}|g\rangle\langle g|\bigg]|\Psi_s\rangle |n\rangle |n\rangle.
\end{eqnarray}
We can look at element by element on the state update,
\begin{eqnarray}
    \text{tr}_{FF'}\{\Pi_o |\Psi_{dt}\rangle\langle \Psi_{dt}|\Pi_0\} &\approx&\sum_n P_n [1-(n+1)\gamma_w dt] |e\rangle\langle e|\rho_s |e\rangle\langle e|\nonumber\\
    &+&\sum_n P_n [1-n\gamma_w dt] |g\rangle\langle g|\rho_s |g\rangle\langle g|+\sum_n P_n [1-(2n+1)\gamma_w dt/2] [|e\rangle\langle e|\rho_s |g\rangle\langle g|+|g\rangle\langle g|\rho_s |e\rangle\langle e|]\nonumber\\
    &=&[1-(\langle n\rangle+1)\gamma_w dt] |e\rangle\langle e|\rho_s |e\rangle\langle e|+(1-\langle n\rangle\gamma_w dt) |g\rangle\langle g|\rho_s |g\rangle\langle g|\nonumber\\
    &+&[1-(2\langle n\rangle+1)\gamma_w dt/2] [|e\rangle\langle e|\rho_s |g\rangle\langle g|+|g\rangle\langle g|\rho_s |e\rangle\langle e|]\nonumber\\
    &\approx&\begin{pmatrix}
         \sqrt{1-(\langle n\rangle +1)\gamma_w dt} &&0\\ 0 && \sqrt{1-\langle n\rangle\gamma_w dt}
     \end{pmatrix}\rho_s\begin{pmatrix}
         \sqrt{1-(\langle n\rangle +1)\gamma_w dt} &&0\\ 0 && \sqrt{1-\langle n\rangle \gamma_w dt}
     \end{pmatrix},\nonumber\\
\end{eqnarray}
where we have used $[1-(2\langle n\rangle+1)\gamma_w dt/2]\approx\sqrt{1-(\langle n\rangle +1)\gamma_w dt}\sqrt{1-\langle n\rangle \gamma_w dt}.$ Thus we obtain the measurement operators corresponding to absorption, emission, and null events as,
\begin{equation}
    \hat{M}_a = -i\begin{pmatrix}
            0  & \sqrt{\langle n\rangle \gamma_w dt}\\
            0  &0
            \end{pmatrix},~~  \hat{M}_e= -i\begin{pmatrix}
            0  & 0\\
            \sqrt{(\langle n\rangle+1)\gamma_w dt}&0
            \end{pmatrix},
\end{equation}
and,
\begin{eqnarray}
    \hat{M}_o = \begin{pmatrix}
         \sqrt{1-(\langle n\rangle +1) \gamma_w dt}&&0\\0&&\sqrt{1-\langle n\rangle \gamma_w dt}
    \end{pmatrix}.
\end{eqnarray}
  
\end{widetext}
\bibliography{ref}

\begin{thebibliography}{61}%
\makeatletter
\providecommand \@ifxundefined [1]{%
 \@ifx{#1\undefined}
}%
\providecommand \@ifnum [1]{%
 \ifnum #1\expandafter \@firstoftwo
 \else \expandafter \@secondoftwo
 \fi
}%
\providecommand \@ifx [1]{%
 \ifx #1\expandafter \@firstoftwo
 \else \expandafter \@secondoftwo
 \fi
}%
\providecommand \natexlab [1]{#1}%
\providecommand \enquote  [1]{``#1''}%
\providecommand \bibnamefont  [1]{#1}%
\providecommand \bibfnamefont [1]{#1}%
\providecommand \citenamefont [1]{#1}%
\providecommand \href@noop [0]{\@secondoftwo}%
\providecommand \href [0]{\begingroup \@sanitize@url \@href}%
\providecommand \@href[1]{\@@startlink{#1}\@@href}%
\providecommand \@@href[1]{\endgroup#1\@@endlink}%
\providecommand \@sanitize@url [0]{\catcode `\\12\catcode `\$12\catcode `\&12\catcode `\#12\catcode `\^12\catcode `\_12\catcode `\%12\relax}%
\providecommand \@@startlink[1]{}%
\providecommand \@@endlink[0]{}%
\providecommand \url  [0]{\begingroup\@sanitize@url \@url }%
\providecommand \@url [1]{\endgroup\@href {#1}{\urlprefix }}%
\providecommand \urlprefix  [0]{URL }%
\providecommand \Eprint [0]{\href }%
\providecommand \doibase [0]{https://doi.org/}%
\providecommand \selectlanguage [0]{\@gobble}%
\providecommand \bibinfo  [0]{\@secondoftwo}%
\providecommand \bibfield  [0]{\@secondoftwo}%
\providecommand \translation [1]{[#1]}%
\providecommand \BibitemOpen [0]{}%
\providecommand \bibitemStop [0]{}%
\providecommand \bibitemNoStop [0]{.\EOS\space}%
\providecommand \EOS [0]{\spacefactor3000\relax}%
\providecommand \BibitemShut  [1]{\csname bibitem#1\endcsname}%
\let\auto@bib@innerbib\@empty
\bibitem [{\citenamefont {Novotny}\ and\ \citenamefont {Hecht}(2012)}]{Novotny_Hecht_2012}%
  \BibitemOpen
  \bibfield  {author} {\bibinfo {author} {\bibfnamefont {L.}~\bibnamefont {Novotny}}\ and\ \bibinfo {author} {\bibfnamefont {B.}~\bibnamefont {Hecht}},\ }\bibinfo {title} {Quantum emitters},\ in\ \href@noop {} {\emph {\bibinfo {booktitle} {Principles of Nano-Optics}}}\ (\bibinfo  {publisher} {Cambridge University Press},\ \bibinfo {year} {2012})\ pp.\ \bibinfo {pages} {282--312}\BibitemShut {NoStop}%
\bibitem [{\citenamefont {Mandel}\ and\ \citenamefont {Wolf}(1995)}]{Mandel_Wolf_1995}%
  \BibitemOpen
  \bibfield  {author} {\bibinfo {author} {\bibfnamefont {L.}~\bibnamefont {Mandel}}\ and\ \bibinfo {author} {\bibfnamefont {E.}~\bibnamefont {Wolf}},\ }\href@noop {} {\emph {\bibinfo {title} {Optical Coherence and Quantum Optics}}}\ (\bibinfo  {publisher} {Cambridge University Press},\ \bibinfo {year} {1995})\BibitemShut {NoStop}%
\bibitem [{\citenamefont {Fortsch}\ \emph {et~al.}(2013)\citenamefont {Fortsch}, \citenamefont {Furst}, \citenamefont {Wittmann}, \citenamefont {Strekalov}, \citenamefont {Aiello}, \citenamefont {Chekhova}, \citenamefont {Silberhorn}, \citenamefont {Leuchs},\ and\ \citenamefont {Marquardt}}]{fortsch_versatile_2013}%
  \BibitemOpen
  \bibfield  {author} {\bibinfo {author} {\bibfnamefont {M.}~\bibnamefont {Fortsch}}, \bibinfo {author} {\bibfnamefont {J.~U.}\ \bibnamefont {Furst}}, \bibinfo {author} {\bibfnamefont {C.}~\bibnamefont {Wittmann}}, \bibinfo {author} {\bibfnamefont {D.}~\bibnamefont {Strekalov}}, \bibinfo {author} {\bibfnamefont {A.}~\bibnamefont {Aiello}}, \bibinfo {author} {\bibfnamefont {M.~V.}\ \bibnamefont {Chekhova}}, \bibinfo {author} {\bibfnamefont {C.}~\bibnamefont {Silberhorn}}, \bibinfo {author} {\bibfnamefont {G.}~\bibnamefont {Leuchs}},\ and\ \bibinfo {author} {\bibfnamefont {C.}~\bibnamefont {Marquardt}},\ }\bibfield  {title} {\bibinfo {title} {A versatile source of single photons for quantum information processing},\ }\href {https://doi.org/10.1038/ncomms2838} {\bibfield  {journal} {\bibinfo  {journal} {Nature Communications}\ }\textbf {\bibinfo {volume} {4}},\ \bibinfo {pages} {1818} (\bibinfo {year} {2013})}\BibitemShut {NoStop}%
\bibitem [{\citenamefont {Yang}\ \emph {et~al.}(2024)\citenamefont {Yang}, \citenamefont {Khanahmadi}, \citenamefont {Strandberg}, \citenamefont {Gaikwad}, \citenamefont {Castillo-Moreno}, \citenamefont {Kockum}, \citenamefont {Ullah}, \citenamefont {Johansson}, \citenamefont {Eriksson},\ and\ \citenamefont {Gasparinetti}}]{yang_deterministic_2024}%
  \BibitemOpen
  \bibfield  {author} {\bibinfo {author} {\bibfnamefont {J.}~\bibnamefont {Yang}}, \bibinfo {author} {\bibfnamefont {M.}~\bibnamefont {Khanahmadi}}, \bibinfo {author} {\bibfnamefont {I.}~\bibnamefont {Strandberg}}, \bibinfo {author} {\bibfnamefont {A.}~\bibnamefont {Gaikwad}}, \bibinfo {author} {\bibfnamefont {C.}~\bibnamefont {Castillo-Moreno}}, \bibinfo {author} {\bibfnamefont {A.~F.}\ \bibnamefont {Kockum}}, \bibinfo {author} {\bibfnamefont {M.~A.}\ \bibnamefont {Ullah}}, \bibinfo {author} {\bibfnamefont {G.}~\bibnamefont {Johansson}}, \bibinfo {author} {\bibfnamefont {A.~M.}\ \bibnamefont {Eriksson}},\ and\ \bibinfo {author} {\bibfnamefont {S.}~\bibnamefont {Gasparinetti}},\ }\href {https://doi.org/10.48550/arXiv.2410.23202} {\bibinfo {title} {Deterministic generation of frequency-bin-encoded microwave photons}} (\bibinfo {year} {2024}),\ \bibinfo {note} {arXiv:2410.23202}\BibitemShut {NoStop}%
\bibitem [{\citenamefont {Castillo-Moreno}\ \emph {et~al.}(2025)\citenamefont {Castillo-Moreno}, \citenamefont {Amin}, \citenamefont {Strandberg}, \citenamefont {Kervinen}, \citenamefont {Osman},\ and\ \citenamefont {Gasparinetti}}]{castillo-moreno_dynamical_2024}%
  \BibitemOpen
  \bibfield  {author} {\bibinfo {author} {\bibfnamefont {C.}~\bibnamefont {Castillo-Moreno}}, \bibinfo {author} {\bibfnamefont {K.~R.}\ \bibnamefont {Amin}}, \bibinfo {author} {\bibfnamefont {I.}~\bibnamefont {Strandberg}}, \bibinfo {author} {\bibfnamefont {M.}~\bibnamefont {Kervinen}}, \bibinfo {author} {\bibfnamefont {A.}~\bibnamefont {Osman}},\ and\ \bibinfo {author} {\bibfnamefont {S.}~\bibnamefont {Gasparinetti}},\ }\bibfield  {title} {\bibinfo {title} {Dynamical excitation control and multimode emission of an atom-photon bound state},\ }\href {https://doi.org/10.1103/PhysRevLett.134.133601} {\bibfield  {journal} {\bibinfo  {journal} {Phys. Rev. Lett.}\ }\textbf {\bibinfo {volume} {134}},\ \bibinfo {pages} {133601} (\bibinfo {year} {2025})}\BibitemShut {NoStop}%
\bibitem [{\citenamefont {Mandel}(1979)}]{mandel_sub-poissonian_1979}%
  \BibitemOpen
  \bibfield  {author} {\bibinfo {author} {\bibfnamefont {L.}~\bibnamefont {Mandel}},\ }\bibfield  {title} {\bibinfo {title} {Sub-{Poissonian} photon statistics in resonance fluorescence},\ }\href {https://doi.org/10.1364/OL.4.000205} {\bibfield  {journal} {\bibinfo  {journal} {Optics Letters}\ }\textbf {\bibinfo {volume} {4}},\ \bibinfo {pages} {205} (\bibinfo {year} {1979})}\BibitemShut {NoStop}%
\bibitem [{\citenamefont {Short}\ and\ \citenamefont {Mandel}(1983)}]{short_observation_1983}%
  \BibitemOpen
  \bibfield  {author} {\bibinfo {author} {\bibfnamefont {R.}~\bibnamefont {Short}}\ and\ \bibinfo {author} {\bibfnamefont {L.}~\bibnamefont {Mandel}},\ }\bibfield  {title} {\bibinfo {title} {Observation of {Sub}-{Poissonian} {Photon} {Statistics}},\ }\href {https://doi.org/10.1103/PhysRevLett.51.384} {\bibfield  {journal} {\bibinfo  {journal} {Physical Review Letters}\ }\textbf {\bibinfo {volume} {51}},\ \bibinfo {pages} {384} (\bibinfo {year} {1983})}\BibitemShut {NoStop}%
\bibitem [{\citenamefont {Davidovich}(1996)}]{subP}%
  \BibitemOpen
  \bibfield  {author} {\bibinfo {author} {\bibfnamefont {L.}~\bibnamefont {Davidovich}},\ }\bibfield  {title} {\bibinfo {title} {Sub-poissonian processes in quantum optics},\ }\href {https://doi.org/10.1103/RevModPhys.68.127} {\bibfield  {journal} {\bibinfo  {journal} {Rev. Mod. Phys.}\ }\textbf {\bibinfo {volume} {68}},\ \bibinfo {pages} {127} (\bibinfo {year} {1996})}\BibitemShut {NoStop}%
\bibitem [{\citenamefont {Jordan}\ and\ \citenamefont {Siddiqi}(2024)}]{jordan_quantum_2024}%
  \BibitemOpen
  \bibfield  {author} {\bibinfo {author} {\bibfnamefont {A.~N.}\ \bibnamefont {Jordan}}\ and\ \bibinfo {author} {\bibfnamefont {I.~A.}\ \bibnamefont {Siddiqi}},\ }\href@noop {} {\emph {\bibinfo {title} {Quantum {Measurement}: {Theory} and {Practice}}}}\ (\bibinfo  {publisher} {Cambridge University Press},\ \bibinfo {year} {2024})\BibitemShut {NoStop}%
\bibitem [{\citenamefont {Korotkov}\ and\ \citenamefont {Jordan}(2006)}]{JordanUndoing}%
  \BibitemOpen
  \bibfield  {author} {\bibinfo {author} {\bibfnamefont {A.~N.}\ \bibnamefont {Korotkov}}\ and\ \bibinfo {author} {\bibfnamefont {A.~N.}\ \bibnamefont {Jordan}},\ }\bibfield  {title} {\bibinfo {title} {Undoing a weak quantum measurement of a solid-state qubit},\ }\href {https://doi.org/10.1103/PhysRevLett.97.166805} {\bibfield  {journal} {\bibinfo  {journal} {Phys. Rev. Lett.}\ }\textbf {\bibinfo {volume} {97}},\ \bibinfo {pages} {166805} (\bibinfo {year} {2006})}\BibitemShut {NoStop}%
\bibitem [{\citenamefont {Jordan}\ and\ \citenamefont {Korotkov}(2010)}]{jordan_uncollapsing_2010}%
  \BibitemOpen
  \bibfield  {author} {\bibinfo {author} {\bibfnamefont {A.~N.}\ \bibnamefont {Jordan}}\ and\ \bibinfo {author} {\bibfnamefont {A.~N.}\ \bibnamefont {Korotkov}},\ }\bibfield  {title} {\bibinfo {title} {Uncollapsing the wavefunction by undoing quantum measurements},\ }\href {https://doi.org/10.1080/00107510903385292} {\bibfield  {journal} {\bibinfo  {journal} {Contemporary Physics}\ }\textbf {\bibinfo {volume} {51}},\ \bibinfo {pages} {125} (\bibinfo {year} {2010})}\BibitemShut {NoStop}%
\bibitem [{\citenamefont {Manikandan}\ and\ \citenamefont {Wilczek}(2025)}]{manikandan_detecting_2024}%
  \BibitemOpen
  \bibfield  {author} {\bibinfo {author} {\bibfnamefont {S.~K.}\ \bibnamefont {Manikandan}}\ and\ \bibinfo {author} {\bibfnamefont {F.}~\bibnamefont {Wilczek}},\ }\bibfield  {title} {\bibinfo {title} {Testing the coherent-state description of radiation fields},\ }\href {https://doi.org/10.1103/PhysRevA.111.033705} {\bibfield  {journal} {\bibinfo  {journal} {Phys. Rev. A}\ }\textbf {\bibinfo {volume} {111}},\ \bibinfo {pages} {033705} (\bibinfo {year} {2025})}\BibitemShut {NoStop}%
\bibitem [{\citenamefont {Purcell}(1995)}]{purcell1995spontaneous}%
  \BibitemOpen
  \bibfield  {author} {\bibinfo {author} {\bibfnamefont {E.~M.}\ \bibnamefont {Purcell}},\ }\bibfield  {title} {\bibinfo {title} {Spontaneous emission probabilities at radio frequencies},\ }in\ \href@noop {} {\emph {\bibinfo {booktitle} {Confined Electrons and Photons}}}\ (\bibinfo  {publisher} {Springer},\ \bibinfo {year} {1995})\ pp.\ \bibinfo {pages} {839--839}\BibitemShut {NoStop}%
\bibitem [{\citenamefont {Agarwal}(1974)}]{agarwal_quantum_1974}%
  \BibitemOpen
  \bibfield  {author} {\bibinfo {author} {\bibfnamefont {G.~S.}\ \bibnamefont {Agarwal}},\ }\bibfield  {title} {\bibinfo {title} {Quantum statistical theories of spontaneous emission and their relation to other approaches},\ }in\ \href {https://doi.org/10.1007/BFb0042382} {\emph {\bibinfo {booktitle} {Quantum {Optics}}}},\ \bibinfo {editor} {edited by\ \bibinfo {editor} {\bibfnamefont {G.}~\bibnamefont {Hohler}}}\ (\bibinfo  {publisher} {Springer},\ \bibinfo {address} {Berlin},\ \bibinfo {year} {1974})\ pp.\ \bibinfo {pages} {1--128}\BibitemShut {NoStop}%
\bibitem [{\citenamefont {Lewalle}\ \emph {et~al.}(2020)\citenamefont {Lewalle}, \citenamefont {Manikandan}, \citenamefont {Elouard},\ and\ \citenamefont {Jordan}}]{lewalle_measuring_2020}%
  \BibitemOpen
  \bibfield  {author} {\bibinfo {author} {\bibfnamefont {P.}~\bibnamefont {Lewalle}}, \bibinfo {author} {\bibfnamefont {S.~K.}\ \bibnamefont {Manikandan}}, \bibinfo {author} {\bibfnamefont {C.}~\bibnamefont {Elouard}},\ and\ \bibinfo {author} {\bibfnamefont {A.~N.}\ \bibnamefont {Jordan}},\ }\bibfield  {title} {\bibinfo {title} {Measuring fluorescence to track a quantum emitter's state: a theory review},\ }\href {https://doi.org/10.1080/00107514.2020.1747201} {\bibfield  {journal} {\bibinfo  {journal} {Contemporary Physics}\ }\textbf {\bibinfo {volume} {61}},\ \bibinfo {pages} {26} (\bibinfo {year} {2020})}\BibitemShut {NoStop}%
\bibitem [{\citenamefont {Jordan}\ \emph {et~al.}(2016)\citenamefont {Jordan}, \citenamefont {Chantasri}, \citenamefont {Rouchon},\ and\ \citenamefont {Huard}}]{jordan_anatomy_2016}%
  \BibitemOpen
  \bibfield  {author} {\bibinfo {author} {\bibfnamefont {A.~N.}\ \bibnamefont {Jordan}}, \bibinfo {author} {\bibfnamefont {A.}~\bibnamefont {Chantasri}}, \bibinfo {author} {\bibfnamefont {P.}~\bibnamefont {Rouchon}},\ and\ \bibinfo {author} {\bibfnamefont {B.}~\bibnamefont {Huard}},\ }\bibfield  {title} {\bibinfo {title} {Anatomy of fluorescence: quantum trajectory statistics from continuously measuring spontaneous emission},\ }\href {https://doi.org/10.1007/s40509-016-0075-9} {\bibfield  {journal} {\bibinfo  {journal} {Quantum Studies: Mathematics and Foundations}\ }\textbf {\bibinfo {volume} {3}},\ \bibinfo {pages} {237} (\bibinfo {year} {2016})}\BibitemShut {NoStop}%
\bibitem [{\citenamefont {Minev}\ \emph {et~al.}(2019)\citenamefont {Minev}, \citenamefont {Mundhada}, \citenamefont {Shankar}, \citenamefont {Reinhold}, \citenamefont {Gutierrez-Jauregui}, \citenamefont {Schoelkopf}, \citenamefont {Mirrahimi}, \citenamefont {Carmichael},\ and\ \citenamefont {Devoret}}]{minev_catch_2019}%
  \BibitemOpen
  \bibfield  {author} {\bibinfo {author} {\bibfnamefont {Z.~K.}\ \bibnamefont {Minev}}, \bibinfo {author} {\bibfnamefont {S.~O.}\ \bibnamefont {Mundhada}}, \bibinfo {author} {\bibfnamefont {S.}~\bibnamefont {Shankar}}, \bibinfo {author} {\bibfnamefont {P.}~\bibnamefont {Reinhold}}, \bibinfo {author} {\bibfnamefont {R.}~\bibnamefont {Gutierrez-Jauregui}}, \bibinfo {author} {\bibfnamefont {R.~J.}\ \bibnamefont {Schoelkopf}}, \bibinfo {author} {\bibfnamefont {M.}~\bibnamefont {Mirrahimi}}, \bibinfo {author} {\bibfnamefont {H.~J.}\ \bibnamefont {Carmichael}},\ and\ \bibinfo {author} {\bibfnamefont {M.~H.}\ \bibnamefont {Devoret}},\ }\bibfield  {title} {\bibinfo {title} {To catch and reverse a quantum jump mid-flight},\ }\href {https://doi.org/10.1038/s41586-019-1287-z} {\bibfield  {journal} {\bibinfo  {journal} {Nature}\ }\textbf {\bibinfo {volume} {570}},\ \bibinfo {pages} {200} (\bibinfo {year} {2019})}\BibitemShut {NoStop}%
\bibitem [{\citenamefont {Manikandan}\ \emph {et~al.}(2019)\citenamefont {Manikandan}, \citenamefont {Elouard},\ and\ \citenamefont {Jordan}}]{PhysRevA.99.022117}%
  \BibitemOpen
  \bibfield  {author} {\bibinfo {author} {\bibfnamefont {S.~K.}\ \bibnamefont {Manikandan}}, \bibinfo {author} {\bibfnamefont {C.}~\bibnamefont {Elouard}},\ and\ \bibinfo {author} {\bibfnamefont {A.~N.}\ \bibnamefont {Jordan}},\ }\bibfield  {title} {\bibinfo {title} {Fluctuation theorems for continuous quantum measurements and absolute irreversibility},\ }\href {https://doi.org/10.1103/PhysRevA.99.022117} {\bibfield  {journal} {\bibinfo  {journal} {Phys. Rev. A}\ }\textbf {\bibinfo {volume} {99}},\ \bibinfo {pages} {022117} (\bibinfo {year} {2019})}\BibitemShut {NoStop}%
\bibitem [{\citenamefont {Kannan}\ \emph {et~al.}(2020)\citenamefont {Kannan}, \citenamefont {Ruckriegel}, \citenamefont {Campbell}, \citenamefont {Frisk~Kockum}, \citenamefont {Braumüller}, \citenamefont {Kim}, \citenamefont {Kjaergaard}, \citenamefont {Krantz}, \citenamefont {Melville}, \citenamefont {Niedzielski}, \citenamefont {Vepsäläinen}, \citenamefont {Winik}, \citenamefont {Yoder}, \citenamefont {Nori}, \citenamefont {Orlando}, \citenamefont {Gustavsson},\ and\ \citenamefont {Oliver}}]{kannan_waveguide_2020}%
  \BibitemOpen
  \bibfield  {author} {\bibinfo {author} {\bibfnamefont {B.}~\bibnamefont {Kannan}}, \bibinfo {author} {\bibfnamefont {M.~J.}\ \bibnamefont {Ruckriegel}}, \bibinfo {author} {\bibfnamefont {D.~L.}\ \bibnamefont {Campbell}}, \bibinfo {author} {\bibfnamefont {A.}~\bibnamefont {Frisk~Kockum}}, \bibinfo {author} {\bibfnamefont {J.}~\bibnamefont {Braumüller}}, \bibinfo {author} {\bibfnamefont {D.~K.}\ \bibnamefont {Kim}}, \bibinfo {author} {\bibfnamefont {M.}~\bibnamefont {Kjaergaard}}, \bibinfo {author} {\bibfnamefont {P.}~\bibnamefont {Krantz}}, \bibinfo {author} {\bibfnamefont {A.}~\bibnamefont {Melville}}, \bibinfo {author} {\bibfnamefont {B.~M.}\ \bibnamefont {Niedzielski}}, \bibinfo {author} {\bibfnamefont {A.}~\bibnamefont {Vepsäläinen}}, \bibinfo {author} {\bibfnamefont {R.}~\bibnamefont {Winik}}, \bibinfo {author} {\bibfnamefont {J.~L.}\ \bibnamefont {Yoder}}, \bibinfo {author} {\bibfnamefont {F.}~\bibnamefont {Nori}}, \bibinfo {author} {\bibfnamefont {T.~P.}\ \bibnamefont {Orlando}}, \bibinfo {author}
  {\bibfnamefont {S.}~\bibnamefont {Gustavsson}},\ and\ \bibinfo {author} {\bibfnamefont {W.~D.}\ \bibnamefont {Oliver}},\ }\bibfield  {title} {\bibinfo {title} {Waveguide quantum electrodynamics with superconducting artificial giant atoms},\ }\href {https://doi.org/10.1038/s41586-020-2529-9} {\bibfield  {journal} {\bibinfo  {journal} {Nature}\ }\textbf {\bibinfo {volume} {583}},\ \bibinfo {pages} {775} (\bibinfo {year} {2020})}\BibitemShut {NoStop}%
\bibitem [{\citenamefont {Wiseman}\ and\ \citenamefont {Brady}(2000)}]{PhysRevA.62.023805}%
  \BibitemOpen
  \bibfield  {author} {\bibinfo {author} {\bibfnamefont {H.~M.}\ \bibnamefont {Wiseman}}\ and\ \bibinfo {author} {\bibfnamefont {Z.}~\bibnamefont {Brady}},\ }\bibfield  {title} {\bibinfo {title} {Robust unravelings for resonance fluorescence},\ }\href {https://doi.org/10.1103/PhysRevA.62.023805} {\bibfield  {journal} {\bibinfo  {journal} {Phys. Rev. A}\ }\textbf {\bibinfo {volume} {62}},\ \bibinfo {pages} {023805} (\bibinfo {year} {2000})}\BibitemShut {NoStop}%
\bibitem [{\citenamefont {Campagne-Ibarcq}\ \emph {et~al.}(2016{\natexlab{a}})\citenamefont {Campagne-Ibarcq}, \citenamefont {Six}, \citenamefont {Bretheau}, \citenamefont {Sarlette}, \citenamefont {Mirrahimi}, \citenamefont {Rouchon},\ and\ \citenamefont {Huard}}]{PhysRevX.6.011002}%
  \BibitemOpen
  \bibfield  {author} {\bibinfo {author} {\bibfnamefont {P.}~\bibnamefont {Campagne-Ibarcq}}, \bibinfo {author} {\bibfnamefont {P.}~\bibnamefont {Six}}, \bibinfo {author} {\bibfnamefont {L.}~\bibnamefont {Bretheau}}, \bibinfo {author} {\bibfnamefont {A.}~\bibnamefont {Sarlette}}, \bibinfo {author} {\bibfnamefont {M.}~\bibnamefont {Mirrahimi}}, \bibinfo {author} {\bibfnamefont {P.}~\bibnamefont {Rouchon}},\ and\ \bibinfo {author} {\bibfnamefont {B.}~\bibnamefont {Huard}},\ }\bibfield  {title} {\bibinfo {title} {Observing quantum state diffusion by heterodyne detection of fluorescence},\ }\href {https://doi.org/10.1103/PhysRevX.6.011002} {\bibfield  {journal} {\bibinfo  {journal} {Phys. Rev. X}\ }\textbf {\bibinfo {volume} {6}},\ \bibinfo {pages} {011002} (\bibinfo {year} {2016}{\natexlab{a}})}\BibitemShut {NoStop}%
\bibitem [{\citenamefont {Ficheux}\ \emph {et~al.}(2018)\citenamefont {Ficheux}, \citenamefont {Jezouin}, \citenamefont {Leghtas},\ and\ \citenamefont {Huard}}]{ficheux_dynamics_2018}%
  \BibitemOpen
  \bibfield  {author} {\bibinfo {author} {\bibfnamefont {Q.}~\bibnamefont {Ficheux}}, \bibinfo {author} {\bibfnamefont {S.}~\bibnamefont {Jezouin}}, \bibinfo {author} {\bibfnamefont {Z.}~\bibnamefont {Leghtas}},\ and\ \bibinfo {author} {\bibfnamefont {B.}~\bibnamefont {Huard}},\ }\bibfield  {title} {\bibinfo {title} {Dynamics of a qubit while simultaneously monitoring its relaxation and dephasing},\ }\href {https://doi.org/10.1038/s41467-018-04372-9} {\bibfield  {journal} {\bibinfo  {journal} {Nature Communications}\ }\textbf {\bibinfo {volume} {9}},\ \bibinfo {pages} {1926} (\bibinfo {year} {2018})}\BibitemShut {NoStop}%
\bibitem [{\citenamefont {Joshi}\ \emph {et~al.}(2023)\citenamefont {Joshi}, \citenamefont {Yang},\ and\ \citenamefont {Mirhosseini}}]{PhysRevX.13.021039}%
  \BibitemOpen
  \bibfield  {author} {\bibinfo {author} {\bibfnamefont {C.}~\bibnamefont {Joshi}}, \bibinfo {author} {\bibfnamefont {F.}~\bibnamefont {Yang}},\ and\ \bibinfo {author} {\bibfnamefont {M.}~\bibnamefont {Mirhosseini}},\ }\bibfield  {title} {\bibinfo {title} {Resonance fluorescence of a chiral artificial atom},\ }\href {https://doi.org/10.1103/PhysRevX.13.021039} {\bibfield  {journal} {\bibinfo  {journal} {Phys. Rev. X}\ }\textbf {\bibinfo {volume} {13}},\ \bibinfo {pages} {021039} (\bibinfo {year} {2023})}\BibitemShut {NoStop}%
\bibitem [{\citenamefont {Hutin}\ \emph {et~al.}(2024)\citenamefont {Hutin}, \citenamefont {Essig}, \citenamefont {Assouly}, \citenamefont {Rouchon}, \citenamefont {Bienfait},\ and\ \citenamefont {Huard}}]{PhysRevLett.133.153602}%
  \BibitemOpen
  \bibfield  {author} {\bibinfo {author} {\bibfnamefont {H.}~\bibnamefont {Hutin}}, \bibinfo {author} {\bibfnamefont {A.}~\bibnamefont {Essig}}, \bibinfo {author} {\bibfnamefont {R.}~\bibnamefont {Assouly}}, \bibinfo {author} {\bibfnamefont {P.}~\bibnamefont {Rouchon}}, \bibinfo {author} {\bibfnamefont {A.}~\bibnamefont {Bienfait}},\ and\ \bibinfo {author} {\bibfnamefont {B.}~\bibnamefont {Huard}},\ }\bibfield  {title} {\bibinfo {title} {Monitoring the energy of a cavity by observing the emission of a repeatedly excited qubit},\ }\href {https://doi.org/10.1103/PhysRevLett.133.153602} {\bibfield  {journal} {\bibinfo  {journal} {Phys. Rev. Lett.}\ }\textbf {\bibinfo {volume} {133}},\ \bibinfo {pages} {153602} (\bibinfo {year} {2024})}\BibitemShut {NoStop}%
\bibitem [{\citenamefont {Campagne-Ibarcq}\ \emph {et~al.}(2016{\natexlab{b}})\citenamefont {Campagne-Ibarcq}, \citenamefont {Jezouin}, \citenamefont {Cottet}, \citenamefont {Six}, \citenamefont {Bretheau}, \citenamefont {Mallet}, \citenamefont {Sarlette}, \citenamefont {Rouchon},\ and\ \citenamefont {Huard}}]{PhysRevLett.117.060502}%
  \BibitemOpen
  \bibfield  {author} {\bibinfo {author} {\bibfnamefont {P.}~\bibnamefont {Campagne-Ibarcq}}, \bibinfo {author} {\bibfnamefont {S.}~\bibnamefont {Jezouin}}, \bibinfo {author} {\bibfnamefont {N.}~\bibnamefont {Cottet}}, \bibinfo {author} {\bibfnamefont {P.}~\bibnamefont {Six}}, \bibinfo {author} {\bibfnamefont {L.}~\bibnamefont {Bretheau}}, \bibinfo {author} {\bibfnamefont {F.}~\bibnamefont {Mallet}}, \bibinfo {author} {\bibfnamefont {A.}~\bibnamefont {Sarlette}}, \bibinfo {author} {\bibfnamefont {P.}~\bibnamefont {Rouchon}},\ and\ \bibinfo {author} {\bibfnamefont {B.}~\bibnamefont {Huard}},\ }\bibfield  {title} {\bibinfo {title} {Using spontaneous emission of a qubit as a resource for feedback control},\ }\href {https://doi.org/10.1103/PhysRevLett.117.060502} {\bibfield  {journal} {\bibinfo  {journal} {Phys. Rev. Lett.}\ }\textbf {\bibinfo {volume} {117}},\ \bibinfo {pages} {060502} (\bibinfo {year} {2016}{\natexlab{b}})}\BibitemShut {NoStop}%
\bibitem [{\citenamefont {Essig}\ \emph {et~al.}(2021)\citenamefont {Essig}, \citenamefont {Ficheux}, \citenamefont {Peronnin}, \citenamefont {Cottet}, \citenamefont {Lescanne}, \citenamefont {Sarlette}, \citenamefont {Rouchon}, \citenamefont {Leghtas},\ and\ \citenamefont {Huard}}]{PhysRevX.11.031045}%
  \BibitemOpen
  \bibfield  {author} {\bibinfo {author} {\bibfnamefont {A.}~\bibnamefont {Essig}}, \bibinfo {author} {\bibfnamefont {Q.}~\bibnamefont {Ficheux}}, \bibinfo {author} {\bibfnamefont {T.}~\bibnamefont {Peronnin}}, \bibinfo {author} {\bibfnamefont {N.}~\bibnamefont {Cottet}}, \bibinfo {author} {\bibfnamefont {R.}~\bibnamefont {Lescanne}}, \bibinfo {author} {\bibfnamefont {A.}~\bibnamefont {Sarlette}}, \bibinfo {author} {\bibfnamefont {P.}~\bibnamefont {Rouchon}}, \bibinfo {author} {\bibfnamefont {Z.}~\bibnamefont {Leghtas}},\ and\ \bibinfo {author} {\bibfnamefont {B.}~\bibnamefont {Huard}},\ }\bibfield  {title} {\bibinfo {title} {Multiplexed photon number measurement},\ }\href {https://doi.org/10.1103/PhysRevX.11.031045} {\bibfield  {journal} {\bibinfo  {journal} {Phys. Rev. X}\ }\textbf {\bibinfo {volume} {11}},\ \bibinfo {pages} {031045} (\bibinfo {year} {2021})}\BibitemShut {NoStop}%
\bibitem [{\citenamefont {Manikandan}(2023)}]{manikandan_clocks}%
  \BibitemOpen
  \bibfield  {author} {\bibinfo {author} {\bibfnamefont {S.~K.}\ \bibnamefont {Manikandan}},\ }\bibfield  {title} {\bibinfo {title} {Autonomous quantum clocks using athermal resources},\ }\href {https://doi.org/10.1103/PhysRevResearch.5.043013} {\bibfield  {journal} {\bibinfo  {journal} {Phys. Rev. Res.}\ }\textbf {\bibinfo {volume} {5}},\ \bibinfo {pages} {043013} (\bibinfo {year} {2023})}\BibitemShut {NoStop}%
\bibitem [{Note1()}]{Note1}%
  \BibitemOpen
  \bibinfo {note} {The optical environment can also be thought of as resonant photodetector in the single-mode approximation for the optical field, a framework typically used in the description of continuous measurements~\cite {gross_qubit_2018,jordan_quantum_2024}.}\BibitemShut {Stop}%
\bibitem [{\citenamefont {Carmichael}\ \emph {et~al.}(1987)\citenamefont {Carmichael}, \citenamefont {Lane},\ and\ \citenamefont {Walls}}]{PhysRevLett.58.2539}%
  \BibitemOpen
  \bibfield  {author} {\bibinfo {author} {\bibfnamefont {H.~J.}\ \bibnamefont {Carmichael}}, \bibinfo {author} {\bibfnamefont {A.~S.}\ \bibnamefont {Lane}},\ and\ \bibinfo {author} {\bibfnamefont {D.~F.}\ \bibnamefont {Walls}},\ }\bibfield  {title} {\bibinfo {title} {Resonance fluorescence from an atom in a squeezed vacuum},\ }\href {https://doi.org/10.1103/PhysRevLett.58.2539} {\bibfield  {journal} {\bibinfo  {journal} {Phys. Rev. Lett.}\ }\textbf {\bibinfo {volume} {58}},\ \bibinfo {pages} {2539} (\bibinfo {year} {1987})}\BibitemShut {NoStop}%
\bibitem [{\citenamefont {Toyli}\ \emph {et~al.}(2016)\citenamefont {Toyli}, \citenamefont {Eddins}, \citenamefont {Boutin}, \citenamefont {Puri}, \citenamefont {Hover}, \citenamefont {Bolkhovsky}, \citenamefont {Oliver}, \citenamefont {Blais},\ and\ \citenamefont {Siddiqi}}]{PhysRevX.6.031004}%
  \BibitemOpen
  \bibfield  {author} {\bibinfo {author} {\bibfnamefont {D.~M.}\ \bibnamefont {Toyli}}, \bibinfo {author} {\bibfnamefont {A.~W.}\ \bibnamefont {Eddins}}, \bibinfo {author} {\bibfnamefont {S.}~\bibnamefont {Boutin}}, \bibinfo {author} {\bibfnamefont {S.}~\bibnamefont {Puri}}, \bibinfo {author} {\bibfnamefont {D.}~\bibnamefont {Hover}}, \bibinfo {author} {\bibfnamefont {V.}~\bibnamefont {Bolkhovsky}}, \bibinfo {author} {\bibfnamefont {W.~D.}\ \bibnamefont {Oliver}}, \bibinfo {author} {\bibfnamefont {A.}~\bibnamefont {Blais}},\ and\ \bibinfo {author} {\bibfnamefont {I.}~\bibnamefont {Siddiqi}},\ }\bibfield  {title} {\bibinfo {title} {Resonance fluorescence from an artificial atom in squeezed vacuum},\ }\href {https://doi.org/10.1103/PhysRevX.6.031004} {\bibfield  {journal} {\bibinfo  {journal} {Phys. Rev. X}\ }\textbf {\bibinfo {volume} {6}},\ \bibinfo {pages} {031004} (\bibinfo {year} {2016})}\BibitemShut {NoStop}%
\bibitem [{\citenamefont {Jacobs}\ and\ \citenamefont {Steck}(2006)}]{jacobs_straightforward_2006}%
  \BibitemOpen
  \bibfield  {author} {\bibinfo {author} {\bibfnamefont {K.}~\bibnamefont {Jacobs}}\ and\ \bibinfo {author} {\bibfnamefont {D.~A.}\ \bibnamefont {Steck}},\ }\bibfield  {title} {\bibinfo {title} {A straightforward introduction to continuous quantum measurement},\ }\href {https://doi.org/10.1080/00107510601101934} {\bibfield  {journal} {\bibinfo  {journal} {Contemporary Physics}\ }\textbf {\bibinfo {volume} {47}},\ \bibinfo {pages} {279} (\bibinfo {year} {2006})}\BibitemShut {NoStop}%
\bibitem [{\citenamefont {Pilgram}\ \emph {et~al.}(2003)\citenamefont {Pilgram}, \citenamefont {Jordan}, \citenamefont {Sukhorukov},\ and\ \citenamefont {B\"uttiker}}]{PhysRevLett.90.206801}%
  \BibitemOpen
  \bibfield  {author} {\bibinfo {author} {\bibfnamefont {S.}~\bibnamefont {Pilgram}}, \bibinfo {author} {\bibfnamefont {A.~N.}\ \bibnamefont {Jordan}}, \bibinfo {author} {\bibfnamefont {E.~V.}\ \bibnamefont {Sukhorukov}},\ and\ \bibinfo {author} {\bibfnamefont {M.}~\bibnamefont {B\"uttiker}},\ }\bibfield  {title} {\bibinfo {title} {Stochastic path integral formulation of full counting statistics},\ }\href {https://doi.org/10.1103/PhysRevLett.90.206801} {\bibfield  {journal} {\bibinfo  {journal} {Phys. Rev. Lett.}\ }\textbf {\bibinfo {volume} {90}},\ \bibinfo {pages} {206801} (\bibinfo {year} {2003})}\BibitemShut {NoStop}%
\bibitem [{\citenamefont {Sukhorukov}\ \emph {et~al.}(2007)\citenamefont {Sukhorukov}, \citenamefont {Jordan}, \citenamefont {Gustavsson}, \citenamefont {Leturcq}, \citenamefont {Ihn},\ and\ \citenamefont {Ensslin}}]{sukhorukov_conditional_2007}%
  \BibitemOpen
  \bibfield  {author} {\bibinfo {author} {\bibfnamefont {E.~V.}\ \bibnamefont {Sukhorukov}}, \bibinfo {author} {\bibfnamefont {A.~N.}\ \bibnamefont {Jordan}}, \bibinfo {author} {\bibfnamefont {S.}~\bibnamefont {Gustavsson}}, \bibinfo {author} {\bibfnamefont {R.}~\bibnamefont {Leturcq}}, \bibinfo {author} {\bibfnamefont {T.}~\bibnamefont {Ihn}},\ and\ \bibinfo {author} {\bibfnamefont {K.}~\bibnamefont {Ensslin}},\ }\bibfield  {title} {\bibinfo {title} {Conditional statistics of electron transport in interacting nanoscale conductors},\ }\href {https://doi.org/10.1038/nphys564} {\bibfield  {journal} {\bibinfo  {journal} {Nature Physics}\ }\textbf {\bibinfo {volume} {3}},\ \bibinfo {pages} {243} (\bibinfo {year} {2007})}\BibitemShut {NoStop}%
\bibitem [{\citenamefont {Bagrets}\ and\ \citenamefont {Nazarov}(2003)}]{PhysRevB.67.085316}%
  \BibitemOpen
  \bibfield  {author} {\bibinfo {author} {\bibfnamefont {D.~A.}\ \bibnamefont {Bagrets}}\ and\ \bibinfo {author} {\bibfnamefont {Y.~V.}\ \bibnamefont {Nazarov}},\ }\bibfield  {title} {\bibinfo {title} {Full counting statistics of charge transfer in coulomb blockade systems},\ }\href {https://doi.org/10.1103/PhysRevB.67.085316} {\bibfield  {journal} {\bibinfo  {journal} {Phys. Rev. B}\ }\textbf {\bibinfo {volume} {67}},\ \bibinfo {pages} {085316} (\bibinfo {year} {2003})}\BibitemShut {NoStop}%
\bibitem [{\citenamefont {Landi}\ \emph {et~al.}(2024)\citenamefont {Landi}, \citenamefont {Kewming}, \citenamefont {Mitchison},\ and\ \citenamefont {Potts}}]{PRXQuantum.5.020201}%
  \BibitemOpen
  \bibfield  {author} {\bibinfo {author} {\bibfnamefont {G.~T.}\ \bibnamefont {Landi}}, \bibinfo {author} {\bibfnamefont {M.~J.}\ \bibnamefont {Kewming}}, \bibinfo {author} {\bibfnamefont {M.~T.}\ \bibnamefont {Mitchison}},\ and\ \bibinfo {author} {\bibfnamefont {P.~P.}\ \bibnamefont {Potts}},\ }\bibfield  {title} {\bibinfo {title} {Current fluctuations in open quantum systems: Bridging the gap between quantum continuous measurements and full counting statistics},\ }\href {https://doi.org/10.1103/PRXQuantum.5.020201} {\bibfield  {journal} {\bibinfo  {journal} {PRX Quantum}\ }\textbf {\bibinfo {volume} {5}},\ \bibinfo {pages} {020201} (\bibinfo {year} {2024})}\BibitemShut {NoStop}%
\bibitem [{\citenamefont {Garrahan}\ and\ \citenamefont {Lesanovsky}(2010)}]{garrahan_thermodynamics_2010}%
  \BibitemOpen
  \bibfield  {author} {\bibinfo {author} {\bibfnamefont {J.~P.}\ \bibnamefont {Garrahan}}\ and\ \bibinfo {author} {\bibfnamefont {I.}~\bibnamefont {Lesanovsky}},\ }\bibfield  {title} {\bibinfo {title} {Thermodynamics of {quantum} {jump} {trajectories}},\ }\href {https://doi.org/10.1103/PhysRevLett.104.160601} {\bibfield  {journal} {\bibinfo  {journal} {Physical Review Letters}\ }\textbf {\bibinfo {volume} {104}},\ \bibinfo {pages} {160601} (\bibinfo {year} {2010})}\BibitemShut {NoStop}%
\bibitem [{\citenamefont {Carollo}\ \emph {et~al.}(2019)\citenamefont {Carollo}, \citenamefont {Jack},\ and\ \citenamefont {Garrahan}}]{PhysRevLett.122.130605}%
  \BibitemOpen
  \bibfield  {author} {\bibinfo {author} {\bibfnamefont {F.}~\bibnamefont {Carollo}}, \bibinfo {author} {\bibfnamefont {R.~L.}\ \bibnamefont {Jack}},\ and\ \bibinfo {author} {\bibfnamefont {J.~P.}\ \bibnamefont {Garrahan}},\ }\bibfield  {title} {\bibinfo {title} {Unraveling the large deviation statistics of markovian open quantum systems},\ }\href {https://doi.org/10.1103/PhysRevLett.122.130605} {\bibfield  {journal} {\bibinfo  {journal} {Phys. Rev. Lett.}\ }\textbf {\bibinfo {volume} {122}},\ \bibinfo {pages} {130605} (\bibinfo {year} {2019})}\BibitemShut {NoStop}%
\bibitem [{\citenamefont {Carollo}\ \emph {et~al.}(2021)\citenamefont {Carollo}, \citenamefont {Garrahan},\ and\ \citenamefont {Jack}}]{carollo_large_2021}%
  \BibitemOpen
  \bibfield  {author} {\bibinfo {author} {\bibfnamefont {F.}~\bibnamefont {Carollo}}, \bibinfo {author} {\bibfnamefont {J.~P.}\ \bibnamefont {Garrahan}},\ and\ \bibinfo {author} {\bibfnamefont {R.~L.}\ \bibnamefont {Jack}},\ }\bibfield  {title} {\bibinfo {title} {Large {Deviations} at {Level} 2.5 for {Markovian} {Open} {Quantum} {Systems}: {Quantum} {Jumps} and {Quantum} {State} {Diffusion}},\ }\href {https://doi.org/10.1007/s10955-021-02799-x} {\bibfield  {journal} {\bibinfo  {journal} {Journal of Statistical Physics}\ }\textbf {\bibinfo {volume} {184}},\ \bibinfo {pages} {13} (\bibinfo {year} {2021})}\BibitemShut {NoStop}%
\bibitem [{\citenamefont {Radaelli}\ \emph {et~al.}(2024)\citenamefont {Radaelli}, \citenamefont {Landi},\ and\ \citenamefont {Binder}}]{Gillespie}%
  \BibitemOpen
  \bibfield  {author} {\bibinfo {author} {\bibfnamefont {M.}~\bibnamefont {Radaelli}}, \bibinfo {author} {\bibfnamefont {G.~T.}\ \bibnamefont {Landi}},\ and\ \bibinfo {author} {\bibfnamefont {F.~C.}\ \bibnamefont {Binder}},\ }\bibfield  {title} {\bibinfo {title} {Gillespie algorithm for quantum jump trajectories},\ }\href {https://doi.org/10.1103/PhysRevA.110.062212} {\bibfield  {journal} {\bibinfo  {journal} {Phys. Rev. A}\ }\textbf {\bibinfo {volume} {110}},\ \bibinfo {pages} {062212} (\bibinfo {year} {2024})}\BibitemShut {NoStop}%
\bibitem [{\citenamefont {Milburn}(2020)}]{milburn_thermodynamics_2020}%
  \BibitemOpen
  \bibfield  {author} {\bibinfo {author} {\bibfnamefont {G.~J.}\ \bibnamefont {Milburn}},\ }\bibfield  {title} {\bibinfo {title} {The thermodynamics of clocks},\ }\href {https://doi.org/10.1080/00107514.2020.1837471} {\bibfield  {journal} {\bibinfo  {journal} {Contemporary Physics}\ }\textbf {\bibinfo {volume} {61}},\ \bibinfo {pages} {69} (\bibinfo {year} {2020})},\ \bibinfo {note} {publisher: Taylor \& Francis \_eprint: https://doi.org/10.1080/00107514.2020.1837471}\BibitemShut {NoStop}%
\bibitem [{\citenamefont {Degen}\ \emph {et~al.}(2017)\citenamefont {Degen}, \citenamefont {Reinhard},\ and\ \citenamefont {Cappellaro}}]{quantumSensing}%
  \BibitemOpen
  \bibfield  {author} {\bibinfo {author} {\bibfnamefont {C.~L.}\ \bibnamefont {Degen}}, \bibinfo {author} {\bibfnamefont {F.}~\bibnamefont {Reinhard}},\ and\ \bibinfo {author} {\bibfnamefont {P.}~\bibnamefont {Cappellaro}},\ }\bibfield  {title} {\bibinfo {title} {Quantum sensing},\ }\href {https://doi.org/10.1103/RevModPhys.89.035002} {\bibfield  {journal} {\bibinfo  {journal} {Rev. Mod. Phys.}\ }\textbf {\bibinfo {volume} {89}},\ \bibinfo {pages} {035002} (\bibinfo {year} {2017})}\BibitemShut {NoStop}%
\bibitem [{\citenamefont {Kim}\ \emph {et~al.}(2002)\citenamefont {Kim}, \citenamefont {Son}, \citenamefont {Bu\ifmmode~\check{z}\else \v{z}\fi{}ek},\ and\ \citenamefont {Knight}}]{PLK}%
  \BibitemOpen
  \bibfield  {author} {\bibinfo {author} {\bibfnamefont {M.~S.}\ \bibnamefont {Kim}}, \bibinfo {author} {\bibfnamefont {W.}~\bibnamefont {Son}}, \bibinfo {author} {\bibfnamefont {V.}~\bibnamefont {Bu\ifmmode~\check{z}\else \v{z}\fi{}ek}},\ and\ \bibinfo {author} {\bibfnamefont {P.~L.}\ \bibnamefont {Knight}},\ }\bibfield  {title} {\bibinfo {title} {Entanglement by a beam splitter: Nonclassicality as a prerequisite for entanglement},\ }\href {https://doi.org/10.1103/PhysRevA.65.032323} {\bibfield  {journal} {\bibinfo  {journal} {Phys. Rev. A}\ }\textbf {\bibinfo {volume} {65}},\ \bibinfo {pages} {032323} (\bibinfo {year} {2002})}\BibitemShut {NoStop}%
\bibitem [{\citenamefont {Eberle}\ \emph {et~al.}(2013)\citenamefont {Eberle}, \citenamefont {Handchen},\ and\ \citenamefont {Schnabel}}]{eberle_stable_2013}%
  \BibitemOpen
  \bibfield  {author} {\bibinfo {author} {\bibfnamefont {T.}~\bibnamefont {Eberle}}, \bibinfo {author} {\bibfnamefont {V.}~\bibnamefont {Handchen}},\ and\ \bibinfo {author} {\bibfnamefont {R.}~\bibnamefont {Schnabel}},\ }\bibfield  {title} {\bibinfo {title} {Stable control of 10 {dB} two-mode squeezed vacuum states of light},\ }\href {https://doi.org/10.1364/OE.21.011546} {\bibfield  {journal} {\bibinfo  {journal} {Optics Express}\ }\textbf {\bibinfo {volume} {21}},\ \bibinfo {pages} {11546} (\bibinfo {year} {2013})}\BibitemShut {NoStop}%
\bibitem [{\citenamefont {Hage}\ \emph {et~al.}(2011)\citenamefont {Hage}, \citenamefont {Janoušek}, \citenamefont {Armstrong}, \citenamefont {Symul}, \citenamefont {Bernu}, \citenamefont {Chrzanowski}, \citenamefont {Lam},\ and\ \citenamefont {Bachor}}]{hage_demonstrating_2011}%
  \BibitemOpen
  \bibfield  {author} {\bibinfo {author} {\bibfnamefont {B.}~\bibnamefont {Hage}}, \bibinfo {author} {\bibfnamefont {J.}~\bibnamefont {Janoušek}}, \bibinfo {author} {\bibfnamefont {S.}~\bibnamefont {Armstrong}}, \bibinfo {author} {\bibfnamefont {T.}~\bibnamefont {Symul}}, \bibinfo {author} {\bibfnamefont {J.}~\bibnamefont {Bernu}}, \bibinfo {author} {\bibfnamefont {H.~M.}\ \bibnamefont {Chrzanowski}}, \bibinfo {author} {\bibfnamefont {P.~K.}\ \bibnamefont {Lam}},\ and\ \bibinfo {author} {\bibfnamefont {H.~A.}\ \bibnamefont {Bachor}},\ }\bibfield  {title} {\bibinfo {title} {Demonstrating various quantum effects with two entangled laser beams},\ }\href {https://doi.org/10.1140/epjd/e2011-20153-9} {\bibfield  {journal} {\bibinfo  {journal} {The European Physical Journal D}\ }\textbf {\bibinfo {volume} {63}},\ \bibinfo {pages} {457} (\bibinfo {year} {2011})}\BibitemShut {NoStop}%
\bibitem [{\citenamefont {Lvovsky}(2015)}]{lvovsky_squeezed_2015}%
  \BibitemOpen
  \bibfield  {author} {\bibinfo {author} {\bibfnamefont {A.~I.}\ \bibnamefont {Lvovsky}},\ }\bibfield  {title} {\bibinfo {title} {Squeezed {Light}},\ }in\ \href {https://doi.org/10.1002/9781119009719.ch5} {\emph {\bibinfo {booktitle} {Photonics}}}\ (\bibinfo  {publisher} {John Wiley \& Sons, New York},\ \bibinfo {year} {2015})\ pp.\ \bibinfo {pages} {121--163}\BibitemShut {NoStop}%
\bibitem [{\citenamefont {Ludlow}\ \emph {et~al.}(2015)\citenamefont {Ludlow}, \citenamefont {Boyd}, \citenamefont {Ye}, \citenamefont {Peik},\ and\ \citenamefont {Schmidt}}]{RevModPhys.87.637}%
  \BibitemOpen
  \bibfield  {author} {\bibinfo {author} {\bibfnamefont {A.~D.}\ \bibnamefont {Ludlow}}, \bibinfo {author} {\bibfnamefont {M.~M.}\ \bibnamefont {Boyd}}, \bibinfo {author} {\bibfnamefont {J.}~\bibnamefont {Ye}}, \bibinfo {author} {\bibfnamefont {E.}~\bibnamefont {Peik}},\ and\ \bibinfo {author} {\bibfnamefont {P.~O.}\ \bibnamefont {Schmidt}},\ }\bibfield  {title} {\bibinfo {title} {Optical atomic clocks},\ }\href {https://doi.org/10.1103/RevModPhys.87.637} {\bibfield  {journal} {\bibinfo  {journal} {Rev. Mod. Phys.}\ }\textbf {\bibinfo {volume} {87}},\ \bibinfo {pages} {637} (\bibinfo {year} {2015})}\BibitemShut {NoStop}%
\bibitem [{\citenamefont {Zhang}\ \emph {et~al.}(2024)\citenamefont {Zhang}, \citenamefont {Ooi}, \citenamefont {Higgins}, \citenamefont {Doyle}, \citenamefont {von~der Wense}, \citenamefont {Beeks}, \citenamefont {Leitner}, \citenamefont {Kazakov}, \citenamefont {Li}, \citenamefont {Thirolf}, \citenamefont {Schumm},\ and\ \citenamefont {Ye}}]{zhang_frequency_2024}%
  \BibitemOpen
  \bibfield  {author} {\bibinfo {author} {\bibfnamefont {C.}~\bibnamefont {Zhang}}, \bibinfo {author} {\bibfnamefont {T.}~\bibnamefont {Ooi}}, \bibinfo {author} {\bibfnamefont {J.~S.}\ \bibnamefont {Higgins}}, \bibinfo {author} {\bibfnamefont {J.~F.}\ \bibnamefont {Doyle}}, \bibinfo {author} {\bibfnamefont {L.}~\bibnamefont {von~der Wense}}, \bibinfo {author} {\bibfnamefont {K.}~\bibnamefont {Beeks}}, \bibinfo {author} {\bibfnamefont {A.}~\bibnamefont {Leitner}}, \bibinfo {author} {\bibfnamefont {G.~A.}\ \bibnamefont {Kazakov}}, \bibinfo {author} {\bibfnamefont {P.}~\bibnamefont {Li}}, \bibinfo {author} {\bibfnamefont {P.~G.}\ \bibnamefont {Thirolf}}, \bibinfo {author} {\bibfnamefont {T.}~\bibnamefont {Schumm}},\ and\ \bibinfo {author} {\bibfnamefont {J.}~\bibnamefont {Ye}},\ }\bibfield  {title} {\bibinfo {title} {Frequency ratio of the {229mTh} nuclear isomeric transition and the {87Sr} atomic clock},\ }\href {https://doi.org/10.1038/s41586-024-07839-6} {\bibfield  {journal} {\bibinfo  {journal} {Nature}\
  }\textbf {\bibinfo {volume} {633}},\ \bibinfo {pages} {63} (\bibinfo {year} {2024})}\BibitemShut {NoStop}%
\bibitem [{\citenamefont {Tiedau}\ \emph {et~al.}(2024)\citenamefont {Tiedau}, \citenamefont {Okhapkin}, \citenamefont {Zhang}, \citenamefont {Thielking}, \citenamefont {Zitzer}, \citenamefont {Peik}, \citenamefont {Schaden}, \citenamefont {Pronebner}, \citenamefont {Morawetz}, \citenamefont {De~Col}, \citenamefont {Schneider}, \citenamefont {Leitner}, \citenamefont {Pressler}, \citenamefont {Kazakov}, \citenamefont {Beeks}, \citenamefont {Sikorsky},\ and\ \citenamefont {Schumm}}]{PhysRevLett.132.182501}%
  \BibitemOpen
  \bibfield  {author} {\bibinfo {author} {\bibfnamefont {J.}~\bibnamefont {Tiedau}}, \bibinfo {author} {\bibfnamefont {M.~V.}\ \bibnamefont {Okhapkin}}, \bibinfo {author} {\bibfnamefont {K.}~\bibnamefont {Zhang}}, \bibinfo {author} {\bibfnamefont {J.}~\bibnamefont {Thielking}}, \bibinfo {author} {\bibfnamefont {G.}~\bibnamefont {Zitzer}}, \bibinfo {author} {\bibfnamefont {E.}~\bibnamefont {Peik}}, \bibinfo {author} {\bibfnamefont {F.}~\bibnamefont {Schaden}}, \bibinfo {author} {\bibfnamefont {T.}~\bibnamefont {Pronebner}}, \bibinfo {author} {\bibfnamefont {I.}~\bibnamefont {Morawetz}}, \bibinfo {author} {\bibfnamefont {L.~T.}\ \bibnamefont {De~Col}}, \bibinfo {author} {\bibfnamefont {F.}~\bibnamefont {Schneider}}, \bibinfo {author} {\bibfnamefont {A.}~\bibnamefont {Leitner}}, \bibinfo {author} {\bibfnamefont {M.}~\bibnamefont {Pressler}}, \bibinfo {author} {\bibfnamefont {G.~A.}\ \bibnamefont {Kazakov}}, \bibinfo {author} {\bibfnamefont {K.}~\bibnamefont {Beeks}}, \bibinfo {author} {\bibfnamefont
  {T.}~\bibnamefont {Sikorsky}},\ and\ \bibinfo {author} {\bibfnamefont {T.}~\bibnamefont {Schumm}},\ }\bibfield  {title} {\bibinfo {title} {Laser excitation of the th-229 nucleus},\ }\href {https://doi.org/10.1103/PhysRevLett.132.182501} {\bibfield  {journal} {\bibinfo  {journal} {Phys. Rev. Lett.}\ }\textbf {\bibinfo {volume} {132}},\ \bibinfo {pages} {182501} (\bibinfo {year} {2024})}\BibitemShut {NoStop}%
\bibitem [{\citenamefont {Panov}(1996)}]{panov_quantum_1996}%
  \BibitemOpen
  \bibfield  {author} {\bibinfo {author} {\bibfnamefont {A.~D.}\ \bibnamefont {Panov}},\ }\bibfield  {title} {\bibinfo {title} {Quantum {Zeno} {Effect}, {Nuclear} {Conversion} and {Photoionization} in {Solids}},\ }\href {https://doi.org/10.1006/aphy.1996.0063} {\bibfield  {journal} {\bibinfo  {journal} {Annals of Physics}\ }\textbf {\bibinfo {volume} {249}},\ \bibinfo {pages} {1} (\bibinfo {year} {1996})}\BibitemShut {NoStop}%
\bibitem [{\citenamefont {He}\ \emph {et~al.}(2023)\citenamefont {He}, \citenamefont {Pakkiam}, \citenamefont {Gangat}, \citenamefont {Kewming}, \citenamefont {Milburn},\ and\ \citenamefont {Fedorov}}]{prasannaClock}%
  \BibitemOpen
  \bibfield  {author} {\bibinfo {author} {\bibfnamefont {X.}~\bibnamefont {He}}, \bibinfo {author} {\bibfnamefont {P.}~\bibnamefont {Pakkiam}}, \bibinfo {author} {\bibfnamefont {A.~A.}\ \bibnamefont {Gangat}}, \bibinfo {author} {\bibfnamefont {M.~J.}\ \bibnamefont {Kewming}}, \bibinfo {author} {\bibfnamefont {G.~J.}\ \bibnamefont {Milburn}},\ and\ \bibinfo {author} {\bibfnamefont {A.}~\bibnamefont {Fedorov}},\ }\bibfield  {title} {\bibinfo {title} {Effect of measurement backaction on quantum clock precision studied with a superconducting circuit},\ }\href {https://doi.org/10.1103/PhysRevApplied.20.034038} {\bibfield  {journal} {\bibinfo  {journal} {Phys. Rev. Appl.}\ }\textbf {\bibinfo {volume} {20}},\ \bibinfo {pages} {034038} (\bibinfo {year} {2023})}\BibitemShut {NoStop}%
\bibitem [{\citenamefont {Erker}\ \emph {et~al.}(2017)\citenamefont {Erker}, \citenamefont {Mitchison}, \citenamefont {Silva}, \citenamefont {Woods}, \citenamefont {Brunner},\ and\ \citenamefont {Huber}}]{ErkerClocks}%
  \BibitemOpen
  \bibfield  {author} {\bibinfo {author} {\bibfnamefont {P.}~\bibnamefont {Erker}}, \bibinfo {author} {\bibfnamefont {M.~T.}\ \bibnamefont {Mitchison}}, \bibinfo {author} {\bibfnamefont {R.}~\bibnamefont {Silva}}, \bibinfo {author} {\bibfnamefont {M.~P.}\ \bibnamefont {Woods}}, \bibinfo {author} {\bibfnamefont {N.}~\bibnamefont {Brunner}},\ and\ \bibinfo {author} {\bibfnamefont {M.}~\bibnamefont {Huber}},\ }\bibfield  {title} {\bibinfo {title} {Autonomous quantum clocks: Does thermodynamics limit our ability to measure time?},\ }\href {https://doi.org/10.1103/PhysRevX.7.031022} {\bibfield  {journal} {\bibinfo  {journal} {Phys. Rev. X}\ }\textbf {\bibinfo {volume} {7}},\ \bibinfo {pages} {031022} (\bibinfo {year} {2017})}\BibitemShut {NoStop}%
\bibitem [{\citenamefont {Trautmann}\ \emph {et~al.}(2016)\citenamefont {Trautmann}, \citenamefont {Alber},\ and\ \citenamefont {Leuchs}}]{PhysRevA.94.033832}%
  \BibitemOpen
  \bibfield  {author} {\bibinfo {author} {\bibfnamefont {N.}~\bibnamefont {Trautmann}}, \bibinfo {author} {\bibfnamefont {G.}~\bibnamefont {Alber}},\ and\ \bibinfo {author} {\bibfnamefont {G.}~\bibnamefont {Leuchs}},\ }\bibfield  {title} {\bibinfo {title} {Efficient single-photon absorption by a trapped moving atom},\ }\href {https://doi.org/10.1103/PhysRevA.94.033832} {\bibfield  {journal} {\bibinfo  {journal} {Phys. Rev. A}\ }\textbf {\bibinfo {volume} {94}},\ \bibinfo {pages} {033832} (\bibinfo {year} {2016})}\BibitemShut {NoStop}%
\bibitem [{\citenamefont {Stobinska}\ \emph {et~al.}(2009)\citenamefont {Stobinska}, \citenamefont {Alber},\ and\ \citenamefont {Leuchs}}]{stobinska_perfect_2009}%
  \BibitemOpen
  \bibfield  {author} {\bibinfo {author} {\bibfnamefont {M.}~\bibnamefont {Stobinska}}, \bibinfo {author} {\bibfnamefont {G.}~\bibnamefont {Alber}},\ and\ \bibinfo {author} {\bibfnamefont {G.}~\bibnamefont {Leuchs}},\ }\bibfield  {title} {\bibinfo {title} {Perfect excitation of a matter qubit by a single photon in free space},\ }\href {https://doi.org/10.1209/0295-5075/86/14007} {\bibfield  {journal} {\bibinfo  {journal} {Europhysics Letters}\ }\textbf {\bibinfo {volume} {86}},\ \bibinfo {pages} {14007} (\bibinfo {year} {2009})}\BibitemShut {NoStop}%
\bibitem [{\citenamefont {Cangemi}\ \emph {et~al.}(2024)\citenamefont {Cangemi}, \citenamefont {Bhadra},\ and\ \citenamefont {Levy}}]{cangemi_quantum_2023}%
  \BibitemOpen
  \bibfield  {author} {\bibinfo {author} {\bibfnamefont {L.~M.}\ \bibnamefont {Cangemi}}, \bibinfo {author} {\bibfnamefont {C.}~\bibnamefont {Bhadra}},\ and\ \bibinfo {author} {\bibfnamefont {A.}~\bibnamefont {Levy}},\ }\bibfield  {title} {\bibinfo {title} {Quantum engines and refrigerators},\ }\href {https://doi.org/https://doi.org/10.1016/j.physrep.2024.07.001} {\bibfield  {journal} {\bibinfo  {journal} {Physics Reports}\ }\textbf {\bibinfo {volume} {1087}},\ \bibinfo {pages} {1} (\bibinfo {year} {2024})}\BibitemShut {NoStop}%
\bibitem [{\citenamefont {Elouard}\ \emph {et~al.}(2017)\citenamefont {Elouard}, \citenamefont {Herrera-Martí}, \citenamefont {Clusel},\ and\ \citenamefont {Auffeves}}]{elouard_role_2017}%
  \BibitemOpen
  \bibfield  {author} {\bibinfo {author} {\bibfnamefont {C.}~\bibnamefont {Elouard}}, \bibinfo {author} {\bibfnamefont {D.~A.}\ \bibnamefont {Herrera-Martí}}, \bibinfo {author} {\bibfnamefont {M.}~\bibnamefont {Clusel}},\ and\ \bibinfo {author} {\bibfnamefont {A.}~\bibnamefont {Auffeves}},\ }\bibfield  {title} {\bibinfo {title} {The role of quantum measurement in stochastic thermodynamics},\ }\href {https://doi.org/10.1038/s41534-017-0008-4} {\bibfield  {journal} {\bibinfo  {journal} {npj Quantum Information}\ }\textbf {\bibinfo {volume} {3}},\ \bibinfo {pages} {1} (\bibinfo {year} {2017})}\BibitemShut {NoStop}%
\bibitem [{\citenamefont {Ferri-Cort\'es}\ \emph {et~al.}(2025)\citenamefont {Ferri-Cort\'es}, \citenamefont {Almanza-Marrero}, \citenamefont {L\'opez}, \citenamefont {Zambrini},\ and\ \citenamefont {Manzano}}]{ferri-cortes_entropy_2024}%
  \BibitemOpen
  \bibfield  {author} {\bibinfo {author} {\bibfnamefont {M.}~\bibnamefont {Ferri-Cort\'es}}, \bibinfo {author} {\bibfnamefont {J.~A.}\ \bibnamefont {Almanza-Marrero}}, \bibinfo {author} {\bibfnamefont {R.}~\bibnamefont {L\'opez}}, \bibinfo {author} {\bibfnamefont {R.}~\bibnamefont {Zambrini}},\ and\ \bibinfo {author} {\bibfnamefont {G.}~\bibnamefont {Manzano}},\ }\bibfield  {title} {\bibinfo {title} {Conditional fluctuation theorems and entropy production for monitored quantum systems under imperfect detection},\ }\href {https://doi.org/10.1103/PhysRevResearch.7.013077} {\bibfield  {journal} {\bibinfo  {journal} {Phys. Rev. Res.}\ }\textbf {\bibinfo {volume} {7}},\ \bibinfo {pages} {013077} (\bibinfo {year} {2025})}\BibitemShut {NoStop}%
\bibitem [{\citenamefont {Manzano}\ \emph {et~al.}(2018)\citenamefont {Manzano}, \citenamefont {Horowitz},\ and\ \citenamefont {Parrondo}}]{PhysRevX.8.031037}%
  \BibitemOpen
  \bibfield  {author} {\bibinfo {author} {\bibfnamefont {G.}~\bibnamefont {Manzano}}, \bibinfo {author} {\bibfnamefont {J.~M.}\ \bibnamefont {Horowitz}},\ and\ \bibinfo {author} {\bibfnamefont {J.~M.~R.}\ \bibnamefont {Parrondo}},\ }\bibfield  {title} {\bibinfo {title} {Quantum fluctuation theorems for arbitrary environments: Adiabatic and nonadiabatic entropy production},\ }\href {https://doi.org/10.1103/PhysRevX.8.031037} {\bibfield  {journal} {\bibinfo  {journal} {Phys. Rev. X}\ }\textbf {\bibinfo {volume} {8}},\ \bibinfo {pages} {031037} (\bibinfo {year} {2018})}\BibitemShut {NoStop}%
\bibitem [{\citenamefont {Sundelin}\ \emph {et~al.}(2024)\citenamefont {Sundelin}, \citenamefont {Aamir}, \citenamefont {Kulkarni}, \citenamefont {Castillo-Moreno},\ and\ \citenamefont {Gasparinetti}}]{sundelin_quantum_2024}%
  \BibitemOpen
  \bibfield  {author} {\bibinfo {author} {\bibfnamefont {S.}~\bibnamefont {Sundelin}}, \bibinfo {author} {\bibfnamefont {M.~A.}\ \bibnamefont {Aamir}}, \bibinfo {author} {\bibfnamefont {V.~M.}\ \bibnamefont {Kulkarni}}, \bibinfo {author} {\bibfnamefont {C.}~\bibnamefont {Castillo-Moreno}},\ and\ \bibinfo {author} {\bibfnamefont {S.}~\bibnamefont {Gasparinetti}},\ }\href {https://doi.org/10.48550/arXiv.2403.03373} {\bibinfo {title} {Quantum refrigeration powered by noise in a superconducting circuit}} (\bibinfo {year} {2024}),\ \bibinfo {note} {arXiv:2403.03373}\BibitemShut {NoStop}%
\bibitem [{\citenamefont {A.~M.~Guzman}\ \emph {et~al.}(2024)\citenamefont {A.~M.~Guzman}, \citenamefont {Erker}, \citenamefont {Gasparinetti}, \citenamefont {Huber},\ and\ \citenamefont {Y.~Halpern}}]{guzman_useful_2024}%
  \BibitemOpen
  \bibfield  {author} {\bibinfo {author} {\bibfnamefont {J.}~\bibnamefont {A.~M.~Guzman}}, \bibinfo {author} {\bibfnamefont {P.}~\bibnamefont {Erker}}, \bibinfo {author} {\bibfnamefont {S.}~\bibnamefont {Gasparinetti}}, \bibinfo {author} {\bibfnamefont {M.}~\bibnamefont {Huber}},\ and\ \bibinfo {author} {\bibfnamefont {N.}~\bibnamefont {Y.~Halpern}},\ }\bibfield  {title} {\bibinfo {title} {Useful autonomous quantum machines},\ }\href {https://doi.org/10.1088/1361-6633/ad8803} {\bibfield  {journal} {\bibinfo  {journal} {Reports on Progress in Physics}\ }\textbf {\bibinfo {volume} {87}},\ \bibinfo {pages} {122001} (\bibinfo {year} {2024})}\BibitemShut {NoStop}%
\bibitem [{\citenamefont {Inc.}()}]{Mathematica}%
  \BibitemOpen
  \bibfield  {author} {\bibinfo {author} {\bibfnamefont {W.~R.}\ \bibnamefont {Inc.}},\ }\href@noop {} {\bibinfo {title} {Mathematica, {V}ersion 12.0}},\ \bibinfo {note} {champaign, IL, 2019}\BibitemShut {NoStop}%
\bibitem [{\citenamefont {Gross}\ \emph {et~al.}(2018)\citenamefont {Gross}, \citenamefont {Caves}, \citenamefont {Milburn},\ and\ \citenamefont {Combes}}]{gross_qubit_2018}%
  \BibitemOpen
  \bibfield  {author} {\bibinfo {author} {\bibfnamefont {J.~A.}\ \bibnamefont {Gross}}, \bibinfo {author} {\bibfnamefont {C.~M.}\ \bibnamefont {Caves}}, \bibinfo {author} {\bibfnamefont {G.~J.}\ \bibnamefont {Milburn}},\ and\ \bibinfo {author} {\bibfnamefont {J.}~\bibnamefont {Combes}},\ }\bibfield  {title} {\bibinfo {title} {Qubit models of weak continuous measurements: markovian conditional and open-system dynamics},\ }\href {https://doi.org/10.1088/2058-9565/aaa39f} {\bibfield  {journal} {\bibinfo  {journal} {Quantum Science and Technology}\ }\textbf {\bibinfo {volume} {3}},\ \bibinfo {pages} {024005} (\bibinfo {year} {2018})}\BibitemShut {NoStop}%
\end{thebibliography}%
		   \end{document}